\documentclass[9pt,twocolumn,twoside]{osajnl}

\journal{josab} 

\setboolean{shortarticle}{false}

\usepackage[normalem]{ulem}
\usepackage{amsthm, amsmath, amssymb}       
\usepackage{graphicx}
\usepackage{caption}
\usepackage{subcaption}
\usepackage{float} 
\usepackage{dcolumn}
\usepackage{bm} 
\usepackage{soul} 
\definecolor{forestgreen}{rgb}{0.13, 0.55, 0.13} 
\usepackage{import} 
\usepackage{blindtext} 
\newenvironment{psmallmatrix}
  {\left(\begin{smallmatrix}}
  {\end{smallmatrix}\right)}
  
\usepackage{macros}

\begin{document}

\title{Terahertz lasing conditions of radiative and nonradiative propagating plasmon modes in graphene-coated cylinders}

\author[1]{Leila Prelat} 
\author[2]{Nicol\'as Passarelli} 
\author[3*]{Ra\'ul Bustos-Mar\'un} 
\author[1*]{Ricardo A. Depine} 

\affil[1]{Grupo de Electromagnetismo Aplicado, Departamento de F\'isica, Universidad de Buenos Aires and IFIBA, Consejo Nacional de Investigaciones Cient\'ificas y T\'ecnicas,  Ciudad Universitaria, Pabell\'on I, Buenos Aires 1428, Argentina.}

\affil[2]{School of Physics, University of Sydney, Sydney, New South Wales 2006, Australia.}

\affil[3]{Instituto de F\'isica Enrique Gaviola, Consejo Nacional de Investigaciones Cient\'ificas y T\'ecnicas and Facultad de Ciencias Qu\'imicas, Universidad Nacional de C\'ordoba, Ciudad Universitaria, C\'ordoba 5000, Argentina.}

\date{Compiled \today}

\affil[*]{Corresponding authors: rbustos@famaf.unc.edu.ar, rdep@df.uba.ar}

\begin{abstract}
There is increasing interest in filling the gap of miniaturized terahertz/mid-infrared radiation sources and, particularly, in incorporating these sources into micro/nanophotonic circuits. By using rigorous electromagnetic methods, we investigate the lasing conditions and the  electric-tunability of \rad{radiative and non radiative propagating} surface-plasmon modes in cylinders made of 
\ra{active materials} coated with a graphene layer.
A detailed analysis of the lasing condition of different surface-plasmon modes shows that there is an abrupt change in the gain required when modes become nonradiative. 
Although radiative modes, subject to both radiation and ohmic losses, are expected to require more gain compensation than nonradiative modes, we find that, counterintuitively, gain compensation is greater for nonradiative modes. This is explained in terms \rad{of a change in the distribution of fields that occurs when the character of modes switches from plasmonic to photonic}. Finally, we assess the feasibility of our proposal by using a realist gain medium and showing that a relatively low population inversion is required for the stimulated emission of the studied system.
\end{abstract}



\maketitle

\section{Introduction}
As semiconductor-based electronics approaches its physical limits, new potential solutions are being explored.  In this context, it is believed that nanophotonic plasmon circuits will play important roles in next-generation information technology  \cite{pan2014}.
Propagating surface plasmon polaritons in nanowires, or other forms of waveguides, are usually excited by focusing laser beams on one end of the nanowire, which is clearly an energetically inefficient strategy.
Moreover, the precise positioning of the nanowire inside the focal region can drastically influence the excitation efficiency of the different modes sustained by the plasmonic waveguide \cite{song2017}.
For these applications, it is then desirable that the electromagnetic sources should be within the micro-/nanoscale and that the modes should be generated \textit{in situ} as propagating modes of the waveguides.

Diffraction imposes a lower bound on the focusing of the light and this creates a problem for the handling of electromagnetic radiation at the nanoscale.
The nanoplasmonic counterpart of a laser, the SPASER (by Surface Plasmon Amplification by Stimulated Emission of Radiation), manages to break this limit by emitting (ideally) surface plasmons instead of photons \cite{bergman2003,liu2017,passarelli2016,amendola2017,passarelli2019,natarov2019,jianhua2020}.
Importantly, the generated surface plasmons are not necessarily coupled to the far-field, but they can also be nonradiative (or dark) modes which nonetheless are able to propagate through a waveguide.
Spasers, just like lasers, require active media, or gain materials, which are typically made of dye molecules, semiconductors nanocrystals, or doped dielectrics, where there is a population inversion that sustains the stimulated emission of radiation. This population inversion can be created optically by a pumping laser \cite{noginov2009,oulton2009} or electrically by charge carriers injection into a semiconductor media 
\begin{figure}[htbp]
    \includegraphics[width=0.38\textwidth]{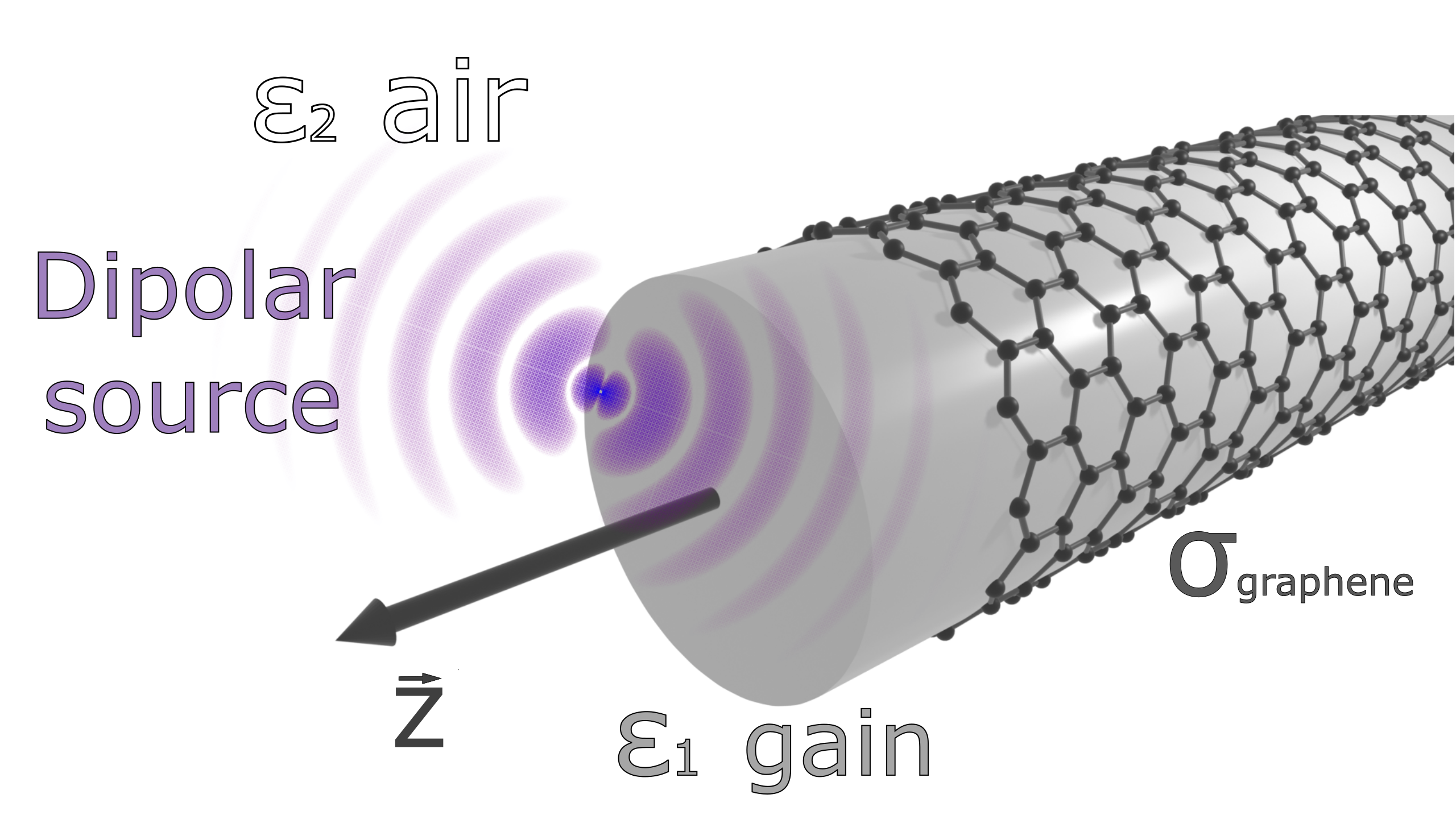}
    \caption{A cylindrical spaser consisting of an active wire coated with a graphene monolayer immersed in an optically transparent medium. We are interested in propagating modes ($k_z \neq 0$). To simulate modal excitation we have used obliquely incident plane waves (to excite radiative modes) and a dipolar source (to excite non-radiative modes). }   \label{dibujo}
\end{figure}

There is a region of the electromagnetic spectrum for which there are few (or none) radiation sources or other practical technological applications. This is known as the ``terahertz gap''. Radiation in this range typically includes electromagnetic waves with frequencies ranging from 0.1 to 10 THz. Due to its ability to preserve samples and the transparency of biological tissues in this frequency range, THz-based technologies have numerous potential applications in biological spectroscopy, detection, and imaging  \cite{falkovsky2008,mittleman2013,kang2018}. 
Clearly, the ability to detect and amplify radiation in this range is highly desirable.
However, it would be even more desirable to incorporate this into nanophotonic circuits, as it  would lead to a new era of lab-on-a-chip medical technologies.

Graphene is an especially interesting material for THz applications due to its transparency, its conductivity controllable with current electronic technologies, and the wide spectral range of excitations it can sustain  \cite{ashcroft1976,geim2005,falkovsky2008,lee2012,weiwei2016,hafez2018,dukhopelnykov2019,cao2022}.
The above reasons have driven the research of graphene plasmons amplification in different structures throughout the literature.
Berman et. al. \cite{berman2013} studied the plasmon amplification of a flat spaser formed by two dense monolayers of quantum dots acting as the active medium and deposited on both sides of a graphene nanoribbon.
Tohari et. al. \cite{tohari2020} proposed a nanospaser consisting of a graphene-metal hybrid plasmon system surrounded by quantum dots.
Ardakani and Faez \cite{ardakani2019} studied a spaser formed by a graphene nanosphere wrapped with two-level quantum dots. 
Azzam et. al. \cite{azzam2020} summarized, up to 2020, the latest advances in spaser theory, while semi-classical models and quantum mechanical formulations for the spaser were explained by Premaratne et. al. \cite{premaratne2017}. 

\rad{Terahertz graphene-based spasers with cylindrical geometries, the kind of structure considered in this paper, were studied in \cite{ardakani2020} and \cite{prelat2021}. Wheras in \cite{ardakani2020} the inner active medium consists of an axially-located single quantum wire and the analysis is limited to the electrostatic regime exclusively, here we consider that the active medium occupies the entire interior and we use a fully retarded rigorous electromagnetic analysis. 
Regarding \cite{prelat2021}, the analyisis is limited to the particular case of localized surface plasmon modes, that is, standing surface waves along the azimuthal direction, with no propagation along the cylinder axis ($k_z=0$, Fig. \ref{dibujo}) and with purely TE polarization (electric field perpendicular to the cylinder axis). By contrast, in this paper we study spasers emitting \textit{propagating} surface plasmon modes ($k_z\neq0$) which are not purely TE, but a combination of both TE and TM (magnetic field perpendicular to the cylinder axis) polarizations. The localized surface plasmons considered in \cite{prelat2021} are radiative, in the sense that they emit radiation towards the exterior medium. Reciprocally, 
they can be coupled to incident radiation without using any additional device, which is why normally incident plane waves were used as the excitation mechanism in Ref. \cite{prelat2021}. 
On the other hand, the propagating modes studied in this paper 
exhibit radiative or nonradiative behavior depending on the value of $k_z$. For $k_z$ values lower than a critical value, the mode is radiative and can be excited by incident radiation, while for $k_z$ values greater than the critical value, the mode does not emit radiation towards the exterior medium and, reciprocally, its excitation requires additional devices (couplers) or near field interactions, such as those produced by nanoantenas. For these  reasons, apart from excitation by a plane wave, in this paper we have also considered excitation by a dipolar source. } 
\ra{These new features make the $k_z\neq0$ modes both richer from a physical point of view and more interesting for applications than the modes with $k_z=0$, particularly in the context of photonic circuits. }

\rad{The main role played by the graphene layer in the system sketched in Fig. \ref{dibujo} is to provide plasma oscillations with characteristics that can be varied by varying its chemical potential. Since we are interested in graphene-based spasers exploiting the tunability of graphene surface plasmons, we want to distinguish plasmonic modes from other type of modes. To do so, we compare the solutions of the rigorous dispersion equation against the solutions of an approximate dispersion equation which extends to $k_z\neq0$ an intuitive plasmonic approach given in \cite{maximo1} for the case $k_z=0$. 
}

The plan of the paper is as follows. We first develop a rigorous electromagnetic method that allows taking advantage of cylindrical geometries to efficiently 
deal with arbitrary $k_z$ values and including the possibility of point-like dipolar exciting fields.
We briefly describe the theory in section \ref{sec:theory}.
Then, we generically study the lasing conditions and the tunability of our system by assuming a wideband approximation for the active medium.
Afterward, we analyze the feasibility of our proposal by performing calculations with a concrete example of a THz active medium taken from recent experiments \cite{pavlov2013,pavlov2020} The main results obtained are presented in section \ref{sec:results} and concluding remarks are provided in Section \ref{sec:conclusions}.

Throughout the manuscript we use Gaussian units and the $\exp(-i\omega t)$ time-dependence is implicit, with $\omega$ the angular frequency, $t$ the time, and $i=\sqrt{-1}$.
The symbols {\rm Re \mit} and {\rm Im \mit} are used for denoting the real and imaginary parts of a complex quantity respectively.

\section{Theory}
\label{sec:theory}

\subsection{Rigorous solution}

The first approach to the problem consisted in solving the Maxwell equations to obtain the electromagnetic fields, applying the boundary conditions and finding the dispersion relation of the modes (plasmonic or not) supported by the system. The modal characteristics depend on the wire size, the constitutive parameters of substrate and ambient media, and the parameters of the graphene surface conductivity. 
Given the cylindrical symmetry of the problem, the longitudinal component of the electromagnetic fields ($E_z$ or $H_z$) can be written as a superposition of discrete partial waves in the form 
\begin{equation} 
F_{n}(\rho,\varphi,z,t)=  F_{n}(\rho) \, \exp{i n \varphi} \, \exp{i(k_z z -\omega_n t)} \,, \label{magnetico}
\end{equation}
with $n$ an integer denoting the mode and $\omega_n$ the complex valued modal frequency. Due to carrier relaxation and radiation losses, plasmon oscillations in passive, dissipative media are always damped and thus the relation 
\begin{equation} 
\Im (\omega_n) < 0 \,, \label{eqn:condomega}
\end{equation}
must be satisfied. Note that this relation holds even when $\Im (\varepsilon_1)=0$, that is, when the wire interior is a completely transparent dielectric medium. This is due to the
losses given by the emission of radiation which are always
present.
The radial dependence $F_{n}(\rho)$ in \eqref{magnetico} can be written as a combination of cylindrical harmonics in the internal ($\rho < R$) and external ($\rho > R$) regions. 
\begin{align} 
\label{Ez12p}
&E_{n}(\rho) =   
\begin{cases} 
   A_{n}\,J_n(k_{t,1}\rho) \,, \text{\, $\rho < R$,}         \\[1mm]
B_{n}\,H_n^{(1)}(k_{t,2}\rho)\,, \text{\, $\rho > R$,}
\end{cases}    
\\[1mm]  
\label{Hz12p}
&H_{n}(\rho) =   
\begin{cases} 
   C_{n}\;J_n(k_{t,1}\rho)  \,, \text{\, $\rho < R$,}         \\[1mm]
 D_{n}\;H_n^{(1)}(k_{t,2}\rho) \,, \text{\, $\rho > R$,}   
\end{cases}      
\end{align}
where $A_{n}, \, B_{n}, \, C_{n}, \, D_{n}$ are unknown complex coefficients, 
$J_n$ and $H_n^{(1)}$ are the $n$-th Bessel and Hankel functions of the first kind respectively, and
\begin{equation}
k_{t,j}=\pm\sqrt{\dfrac{\omega^2}{c^2}\varepsilon_j\mu_j - k_z^2}, \label{eq:k_t}
\end{equation}
where $j=1, 2$ and $c$ is the speed of light in vacuum. The other components of the fields can be obtained from the Ampére’s law and the Faraday-Lenz's law \cite{jackson}
\begin{align*}
& \nabla\times\vec{H} =- \dfrac{i \omega \varepsilon}{c} \vec{E} ,    \\
& \nabla\times \vec{E}=\dfrac{i \omega \mu}{c} \vec{H}    .
\end{align*}
Thus, the transverse part of the electric ($\vec{E}_{t,n}$) and magnetic  ($\vec{H}_{t,n}$) field for the \mbox{$n$-th} mode and the $j$-medium can be written in terms of the $z$-components of the fields \eqref{Ez12p}, \eqref{Hz12p}
\begin{eqnarray}
\vec{E}_{t,n} (\rho)  =   
\Big( \dfrac{-n\omega \mu_j}{c \rho k^2_{t,j}} H_{n,j} + 
\dfrac{ik_z}{k^2_{t,j}} \dfrac{\partial E_{n,j}}{\partial \rho} \Big) \hat{\rho}  \,+  \nonumber \\[1mm] 
 \Big(\dfrac{-n k_z}{\rho k_{t,j}^2} E_{n,j} - \dfrac{i\omega \mu_j}{c k_{t,j}^2} \dfrac{\partial H_{n,j}}{\partial \rho}\Big) \hat{\varphi} \,,
 \label{Et12p}
\end{eqnarray}
\begin{eqnarray}
\vec{H}_{t,n} (\rho)  =   
\Big(\dfrac{i k_z}{k^2_{t,j}}\dfrac{\partial H_{n,j}}{\partial \rho} + 
\dfrac{n\omega \varepsilon_j}{c \rho k^2_{t,j}} E_{n,j} \Big) \hat{\rho}  \,+  \nonumber \\[1mm] 
\Big(\dfrac{-n k_z}{\rho k^2_{t,j}} H_{n,j} + \dfrac{i\omega \varepsilon_j}{c k^2_{t,j}}  \dfrac{\partial E_{n,j}}{\partial \rho}\Big) \hat{\varphi} \, 
\label{Ht12p}
\end{eqnarray}
\rad{where $H_{n,j}$ and $E_{n,j}$ depend on $\rho$ and $\varphi$ as indicated in \eqref{magnetico}. }
\rad{
Using the boundary conditions at the graphene layer (see Appendix \ref{app:boundary}) we obtain a homogeneous system of linear equations 
\begin{equation}
\label{eq:disp01}
\mathbb{M}_n\vec{X}_n = 0\,,
\end{equation}
with the modal amplitudes $\vec{X}_n=\left[A_n, \, B_n, \, C_n, \, D_n \right]^T$ as unknowns. The dispersion relation 
is obtained by finding the non trivial roots of the determinant of $\mathbb{M}_n$. 
The eigenfrequency $\omega_n(k_z)$ fixes the relation between the modal amplitudes of the inner ($A_{n}, \, C_{n}$) and outer ($B_{n}, \, D_{n}$) regions. }
\rad{To distinguish plasmonic modes from other type of modes, the values of the eigenfrequencies $\omega_n(k_z)$ that make zero the determinant of $\mathbb{M}_n$ will be compared against the solutions of the approximate dispersion equation obtained in Appendix \ref{app:aprox}, explicitly  assuming that a surface plasmon is propagating along the graphene cylinder with $k_z\neq0$. }

\rad{In addition to the eigenmode approach, we studied the lasing conditions in terms of scattering and near field observables when the graphene-coated cylinder is excited by monochromatic sources. We have considered two cases: a) an obliquely incident plane wave, and b) an interior dipolar source. In both cases, the solution to the electromagnetic problem follows very symilar steps to those sketched above, except that now the boundary conditions give a non--homogeneous system of linear equations 
\begin{equation}
\label{eq:disp02}
\mathbb{M}_n\vec{X}_n = \vec{V}_n
\end{equation}
with the amplitudes of the excited fields $\vec{X}_n=\left[A_n, \, B_n, \, C_n, \, D_n \right]^T$ as unknowns. The inhomogeneity $\vec{V}_n$ is fixed by the excitation (plane wave or dipole) and $\mathbb{M}_n$ is the same matrix as the homogenous problem \eqref{eq:disp01}. See Appendix \ref{app:boundary} for the expressions of $\mathbb{M}_n$ and $\vec{V}_n$. 
In the case of plane wave excitation, the coefficients $B_{n}, \, D_{n}$ for the outer region can be used to calculate the scattering cross section per unit length of the cylinder 
\begin{align}
\label{eq:Q_scat}
Q_{\text{scat}} =  \dfrac{1}{4\pi R} \dfrac{\omega}{k^2_{t,2}} \sum_n \Big[ \mu_2 |D^{(2)}_n|^2  - \varepsilon_2 |B^{(2)}_n|^2   \Big]. 
\end{align}
}
\subsection{Material modeling}
\nico{Figure \ref{dibujo} presents a view of the nanoscale spaser
based on a graphene--coated cylinder with circular cross--section (radius $R$) centered at $x$=0, $y$=0. The configuration is embedded in a transparent medium with real valued electric permittivity $\varepsilon_{2}$ and magnetic permeability $\mu_{2}$. 
The active core is assumed to have a magnetic permeability $\mu_{1}=1 $ and a complex electric permittivity $\varepsilon_{1}$. We considered two models for the $\varepsilon_{1}$ of the active media: a complex constant for the first one ($\varepsilon_{1} =\Re\varepsilon_1 + i\,\Im (\varepsilon_1)$ with $\Re \varepsilon_1>0$ and $\Im (\varepsilon_1)<0$) and, second, a Lorentz model $\varepsilon_{1}(\omega)$ based on a three-level system.}
\ra{In the first model (wideband approximation), the gain coefficient of the core is $\beta_g=-k_0 \,\Im (\varepsilon_1)/\sqrt{\Re \varepsilon_1}$ \cite{maier2006}, where $k_0=\omega/c$ is the free space propagation constant.
The second model (Lorentz model) accounts for the permittivity of a realistic active medium $\varepsilon$ with a frequency-dependence.
This is written as \cite{passarelli2019,novotny2006}}
\begin{equation}
\label{eq:epsilon_Lorentz}
\varepsilon=\varepsilon_\infty-\sum_{i}\varepsilon_i\chi_i\eta_i
\end{equation}
\nico{where the term $\varepsilon_\infty$ fits the high-frequency behavior, $\eta_i$ is the normalized population difference between the initial and final state of the quantum transition $i$, $\varepsilon_i$ gives the imaginary part of the permittivity of the transition at resonance, and $\chi_i$ is the complex susceptibility, which shapes the spectral profile of the quantum transitions of the material. The latter is usually modeled by a Lorentz-like function
}
\begin{equation}
\label{eq:chi_i}
\chi_i=\frac{\left(\omega_i-\omega+i\gamma_i/2\right)\gamma_i/2}{\left(\omega_i-\omega\right)^2+\left(\gamma_i/2\right)^2}
\end{equation}
\nico{
Resonances occur at an angular frequency $\omega$ equal to $\omega_i$ and have a width $\gamma_i$.
The wide-band approximation discussed before is valid near resonant frequencies ($\omega \approx \omega_i$), where $\varepsilon \approx \varepsilon_\infty - i \varepsilon_i \eta_i $.
}

\begin{figure}[h!]
\centering
\subfloat[$\Re(\epsilon)$]{\includegraphics[width=0.25\textwidth]{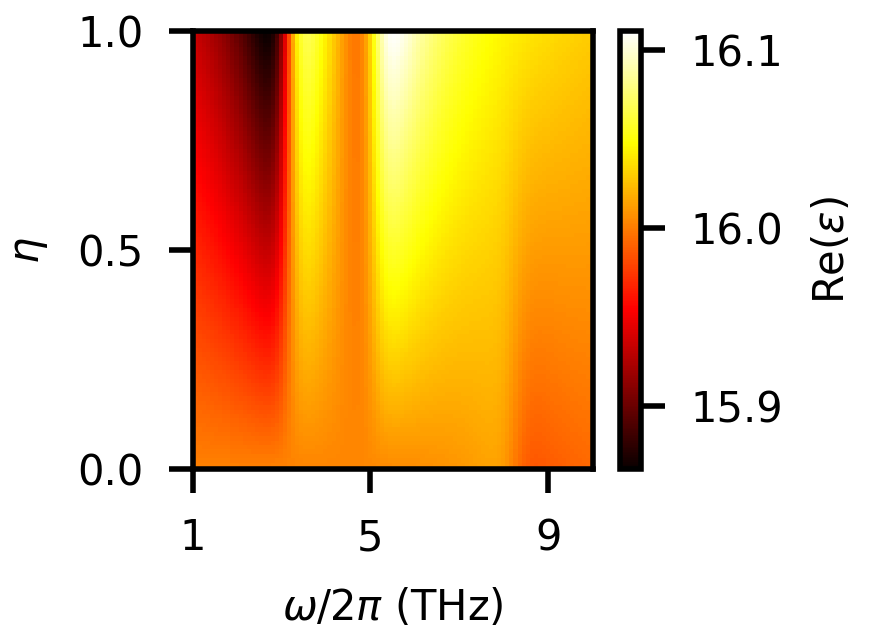}\label{fig5a}}
\subfloat[$\Im(\epsilon)$]{\includegraphics[width=0.25\textwidth]{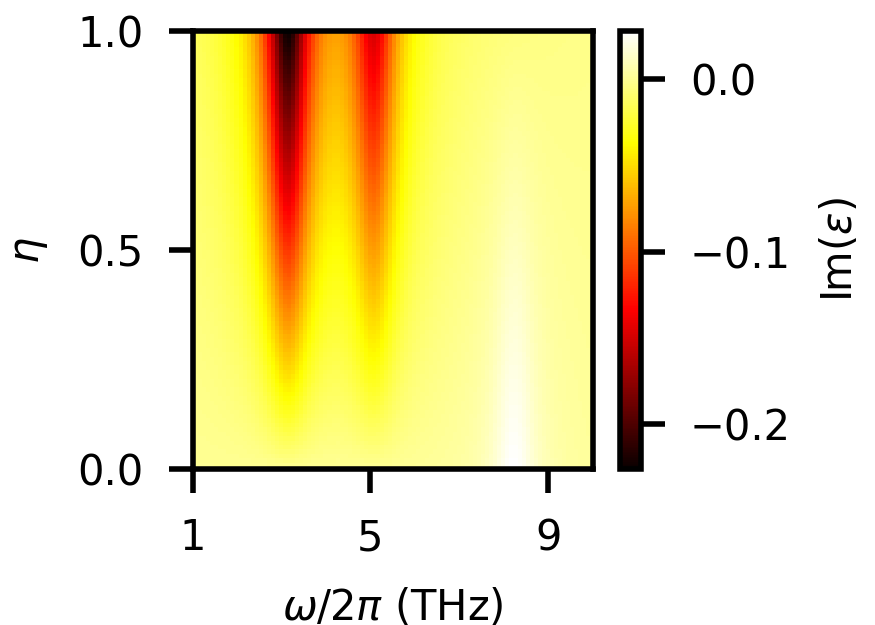}\label{fig5b}}
\caption{(a) real and (b) imaginary part of the permittivity of silicon doped with $N=3\cdot{10}^{15}{cm}^{-3}$ atoms of phosphorus. See the main text for more details.}
\label{fig5}
\end{figure}

\begin{figure}[t]
    \centering
    \subfloat[$k_z = 0.05 \micrometro$]{\includegraphics[width=0.22\textwidth
    ]{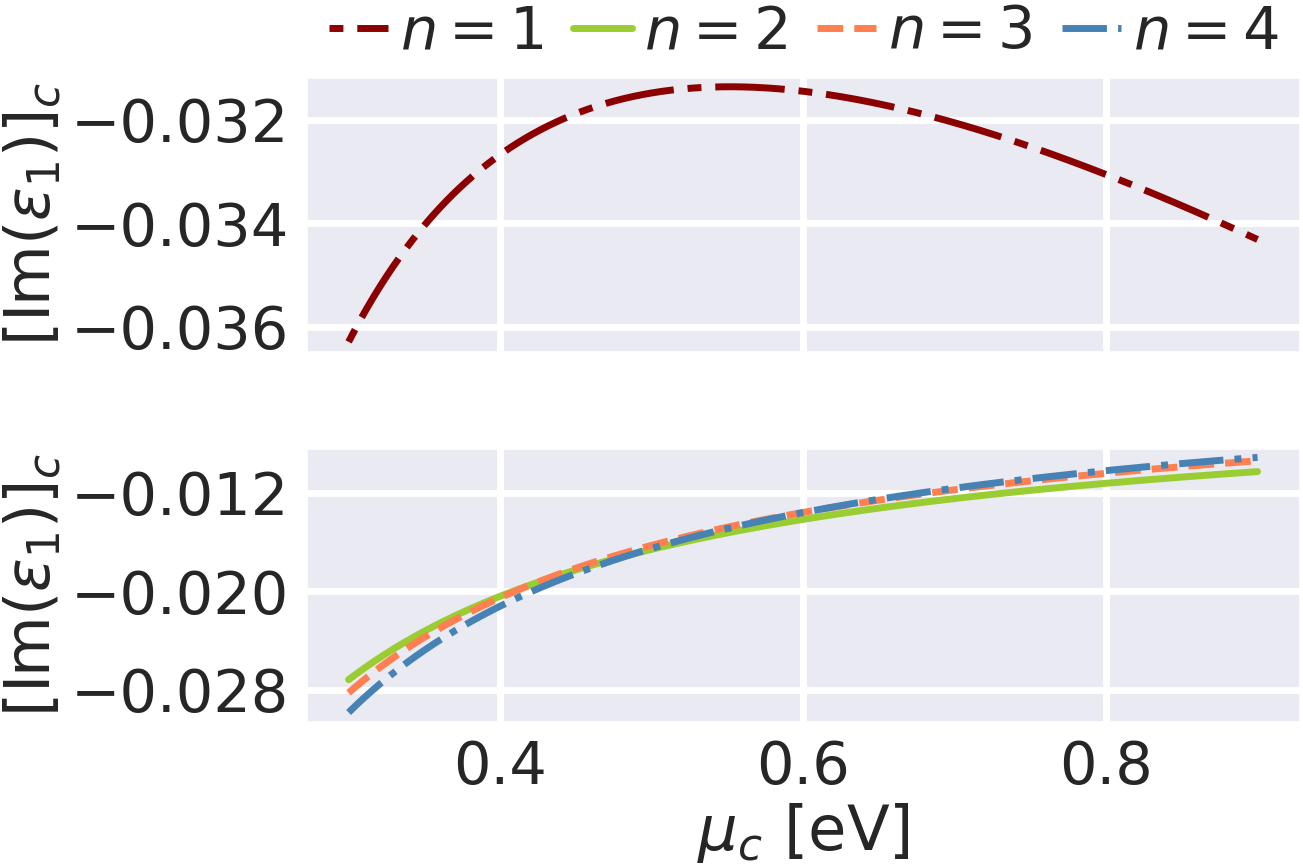}\label{fig3a}}
    \subfloat[$k_z = 0.13 \micrometro$]{\includegraphics[width=0.22\textwidth
    ]{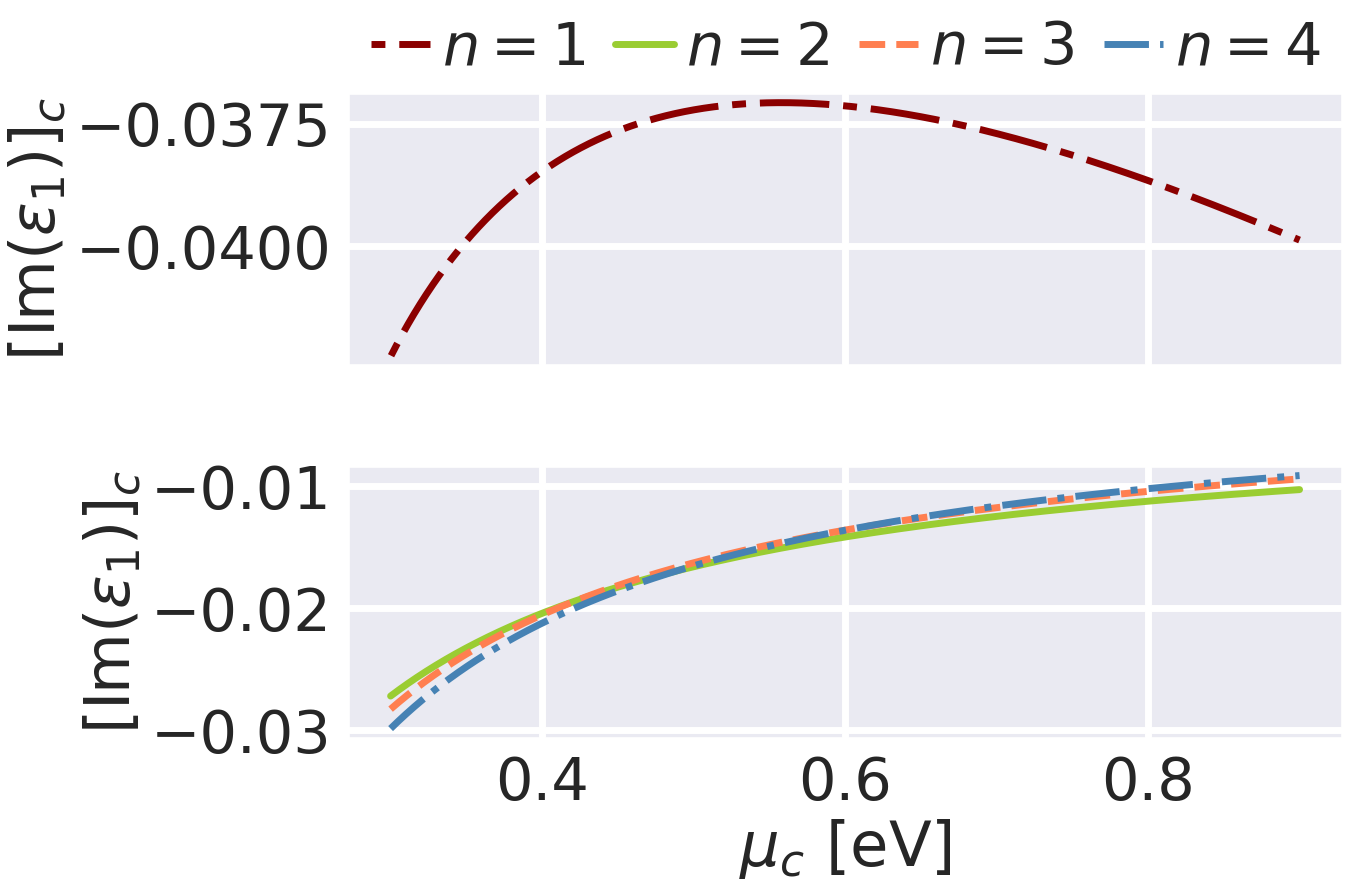}\label{fig3b}}
\\
    \subfloat[$k_z = 0.14 \micrometro$]{\includegraphics[width=0.22\textwidth
    ]{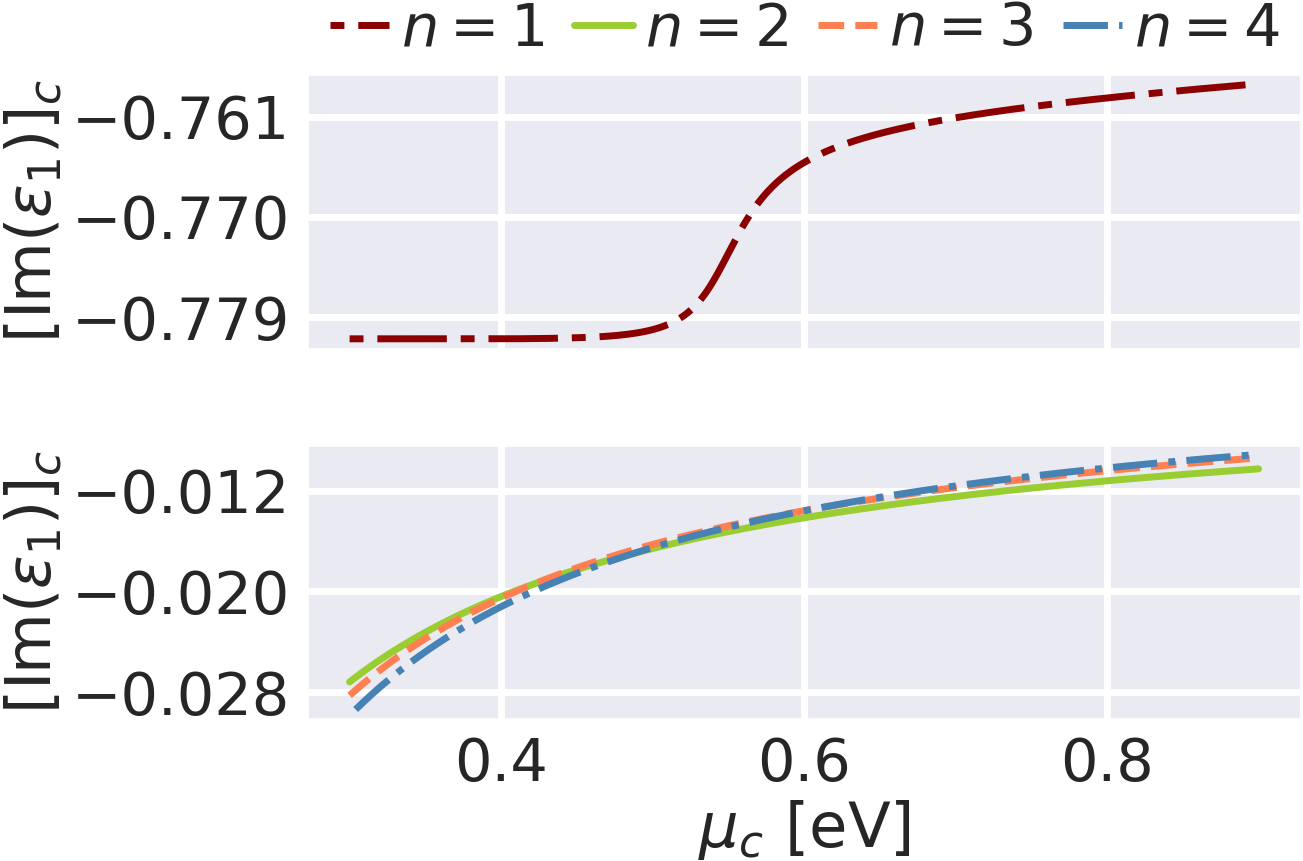}\label{fig3c}}
    \subfloat[$k_z = 0.50 \micrometro$]{\includegraphics[width=0.22\textwidth
    ]{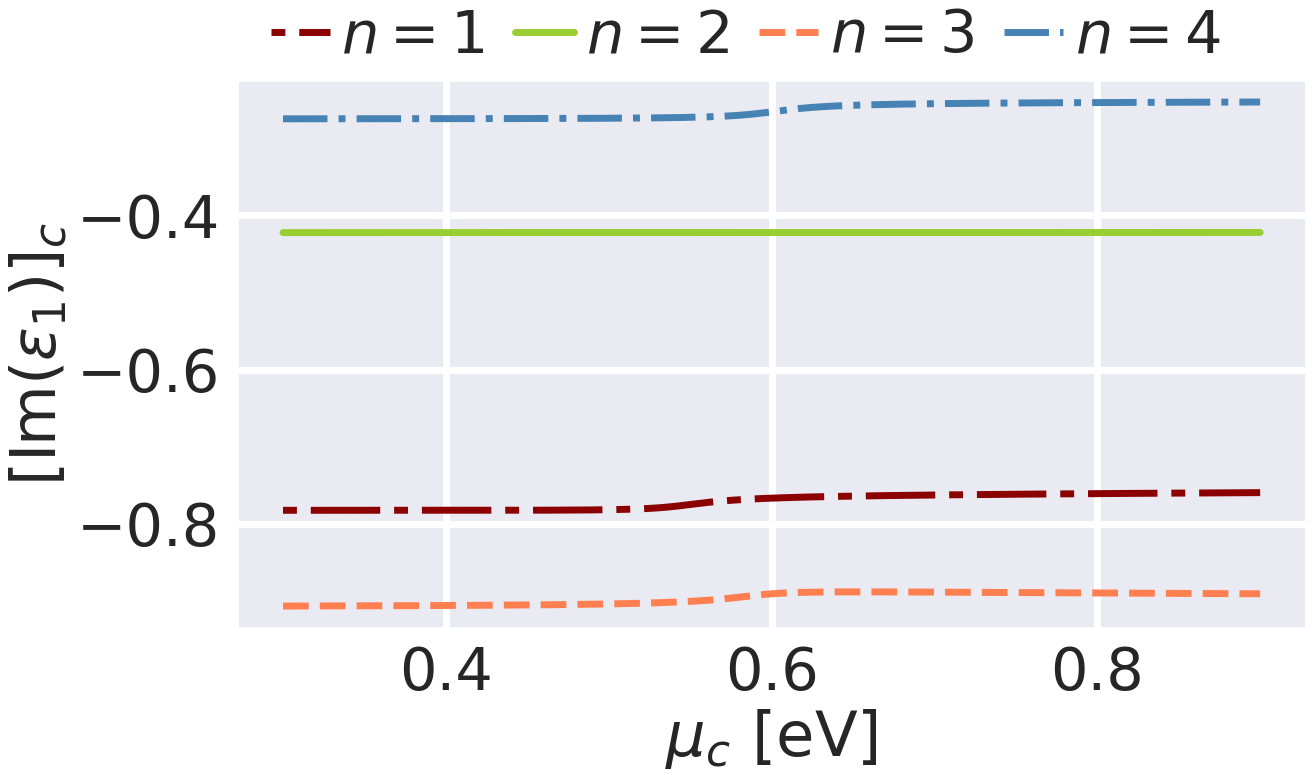}\label{fig3d}}
    \caption{$[\Im (\varepsilon_1)]_c$ as a function of $\mu_c$ for different modes, $n$, and $k_z$ values. The rest of the parameters are \basicTwo}\label{fig3}
\end{figure}

\nico{
For our example in section \ref{sec:results}\ref{sec:realistic} we have chosen phosphorous doped silicon as the gain material. This media was characterized in Ref.  \cite{pavlov2020terahertz}. There, the authors describe the system by three quantum states providing two type of transitions for gain: population inversion (subscript ‘inv’), and stimulated Raman scattering (subscript ‘srs’). These resonances are centered at $5.09$THz and $3.15$THz respectively, while the absorption (subscript ‘abs’) is centered at $8.24$THz.
For this model, Eq. \ref{eq:epsilon_Lorentz} is taken here as: 
\begin{eqnarray}
\label{eq:epsilon_3levels}
\varepsilon_1 & = & \varepsilon_\infty-\varepsilon_{abs}\chi_{abs}\eta-\varepsilon_{inv\ }\ \chi_{inv}\left(1-\eta\right) \notag \\
& & -\varepsilon_{srs\ }\ \chi_{srs}\left(1-\eta\right)
\end{eqnarray}
where the parameter $\eta$, representing an effectively population inversion, will be used as a free parameter.
}
\nico{
According to Ref. \cite{pavlov2020terahertz}, a cross-section of $4\cdot{10}^{-15}{cm}^2$ is expected for a concentration of phosphorous atoms of $N=3\cdot{10}^{15}{cm}^{-3}$ while the value of $\gamma_i$ close to $1THz$ is in accordance with the figures shown in that reference.
For silicon, $\varepsilon_\infty\simeq16$ is the dominant term.
The absorption cross-sections are proportional to the imaginary part of the refractive index $\Im\left(n\right)=\frac{N\sigma_{abs}\lambda_{abs}}{4\pi}$, and thus $\Im \varepsilon_{abs}= 2 \Re \left(n\right) \Im \left(n\right)\simeq 8 \Im\left(n\right)$. The permittivity can be retrieved from the emission cross-section using the F\"uchtbauer-Ladenburg method \cite{passarelli2019}, relating it to the dipolar moment of the transition $\left|\mu_i\right|=\sqrt{\frac{\sigma_i\hbar\varepsilon_\infty\gamma_i}{8\pi nk}}$ through $\varepsilon_i=\frac{8\pi N{|\mu_i|}^2}{\hbar\gamma_i}$.
Fig. \ref{fig5} shows the values of the dielectric function for the above set of parameters.
\footnote{Note that, in order to follow Ref. \cite{pavlov2020terahertz} for the characteristic of the active medium, the value of $\Re (\varepsilon_1)$ becomes $\sim 16$, which is four times larger than the value used in section \ref{sec:results}\ref{sec:general}. Thus, the $k_z$- and $n$-dependent values of $[\omega]_c$ and $[\Im (\varepsilon_1)]_c$ shown in Fig. \ref{fig3} can not be directly compared with Fig. \ref{fig5b}.}
}

The graphene layer is treated as an infinitesimaly thin, local and isotropic two-sided layer with complex valued surface conductivity $\sigma(\omega)$ and we assume that R is large enough, so that the constitutive properties of the graphene coating are the same as those of planar graphene. In this way, we can write 
$\sigma(\omega)=\sigma ^{intra}(\omega) +  \sigma^{inter}(\omega)$, with intraband ($\sigma ^{intra}$) and interband ($\sigma ^{inter}$) transition contributions given by the high frequency expression derived from the Kubo formula (equation (1), Ref.  \cite{kubo1}) 

\begin{equation} \label{intra}
\sigma^{intra}(\omega)= \frac{2i e^2 k_B T}{\pi \hbar (\hbar\omega+i\gamma_c)} \mbox{ln}\left[2 \mbox{cosh}(\mu_c/2 k_B T)\right],
\end{equation}  
\begin{eqnarray} \label{inter}
\sigma^{inter}(\omega)= && 
\frac{e^2}{\hbar} \bigg\{   \frac{1}{2}+\frac{1}{\pi}\mbox{arctan}\left[(\hbar\omega+i\gamma_c-2\mu_c)/2k_BT\right]+ \nonumber \\
&& -  \frac{i}{2\pi}\mbox{ln}\left[\frac{(\hbar\omega+i\gamma_c+2\mu_c)^2}{(\hbar\omega+i\gamma_c-2\mu_c)^2+(2k_BT)^2}\right] \bigg\}\,, \\ \nonumber 
\end{eqnarray}  
where $\mu_c$ is the chemical potential (controlled with the help of a gate voltage), $T$ the ambient temperature, $\gamma_c$ the carriers scattering rate, $k_B$ the Boltzmann constant, $\hbar$ the reduced Planck constant and $e$ the elementary charge. 
For large doping, $\mu_c\ll k_BT$, the intraband contribution (\ref{intra}) dominates and takes the form predicted by the Drude model, whereas the interband contribution (\ref{inter}) dominates for large frequencies $\hbar \omega > \mu_c$ \cite{kubo1,kubo2}. 

\rad{Although we are using the local surface conductivity corresponding to planar graphene, for sufficiently small values of $R$ (sufficiently great curvature), nonlocal effects become important and this aproximation ceases to hold. The validity of the local approximation can be estimated following the arguments given in \cite{CD2021}, where the surface plasmon orbital phase velocity $v_\phi$ is compared with the electron Fermi velocity $v_F \approx  10^{12} \mu m/s$. We have checked that the value of the parameters in our examples verify the limit of applicability of the local conductivity model given by Eq. E3 in  \cite{CD2021}. }

\rad{It should be noted that in some figures of the manuscript, we have shown plots where that range of $\mu_c$ extends up to 1 eV in some cases. These values could be questionable from an experimental point of view, particularly for geometries with a small cylinder radius. However, we do not use the high values of $\mu_c$ to do any claim about the studied configuration, only for theoretical interest, to show the trend of the curves.
}

\subsection{\ra{Lasing condition}}
An \ra{active medium} relies on the stimulated emission of radiation, and that depends, in turn, on the population inversion between an excited and a ground state of the active components of the medium, dyes, quantum dots, rare earth elements, etc.
\ra{For low electromagnetic fields, this population inversion can be taken as constant. However, when the fields are too intense the population inversion as it was can no longer be sustained by the pumping mechanism, and a special treatment is required.
This involves solving self-consistent rate equations which typically consider the population of excited and ground states of the active material, the spatial distribution of electromagnetic fields, spontaneous emission (Purcell-enhanced or not), and stimulated emission \cite{khurgin2020}.
As discussed in several references \cite{arnold2015,passarelli2016}, the consequence of not taking into account the dependence on the field intensity of the materials (linear electrodynamics) is that fields go to infinity once optical losses are exactly compensated.}
This is not a problem in principle if one is only interested in finding lasing conditions and not the intensity of the fields. Indeed, this property can be used for finding 
\ra{the gain-loss compensation condition} as divergences of properties such as the scattering coefficient \cite{passarelli2016,passarelli2019}.
Another strategy to study lasing conditions is to find the critical value of the imaginary part of $\varepsilon_1$ for which the modal eigenfrequency $\omega_n$ is real \cite{smotrova2011,natarov2019,prelat2021}.
Here we use both strategies\ra{, but always within linear electrodynamics.}
The true intensity of electromagnetic fields at the lasing condition is beyond the scope of this work.

\ra{At this pont a clarification is in order. Some authors define lasing threshold as the the condition when the stimulated emission rate equals the spontaneous emission rate  \cite{khurgin2020}.
This requires a gain greater than that provided by the gain-loss compensation condition. Despite this, our lasing condition is helpful to estimate the minimum gain required to observe an important enhancement of near and far fields. Moreover, our results will be helpful to correctly predict the mode that will be lasing (or spasing) and the frequency at which this will occur.}
\begin{figure}[t]
    \centering
    \subfloat[$k_z = 0.13 \micrometro$]{\includegraphics[width=0.22\textwidth
    ]{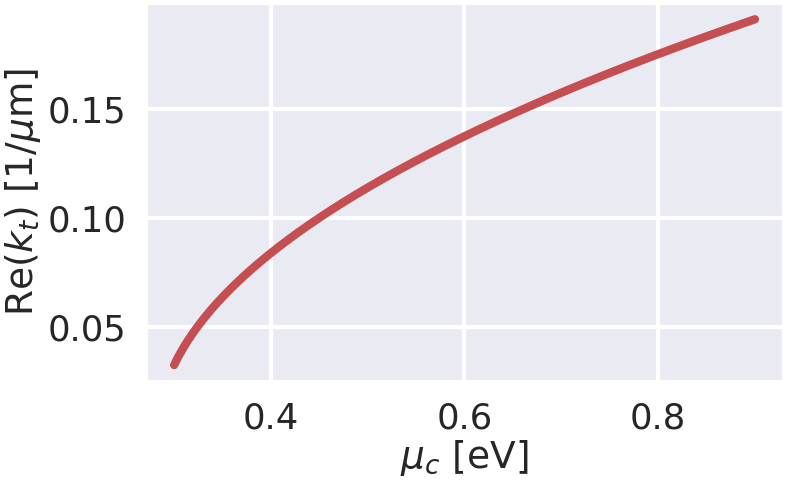}\label{fig35a}}
    \subfloat[$k_z = 0.14 \micrometro$]{\includegraphics[width=0.22\textwidth
    ]{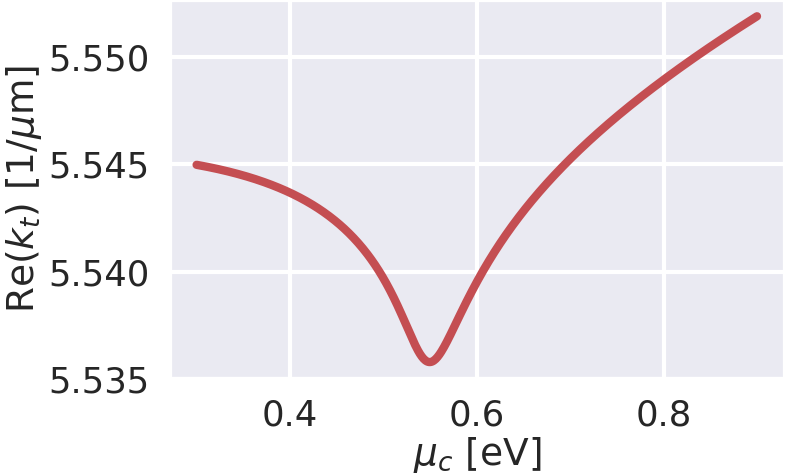}\label{fig35b}}
    \caption{Value of $k_{t,2}$ vs $\mu$ for the lasing condition of mode $n=1$ for two different values of $k_z$. The rest of the parameters are the same as in fig. \ref{fig3}. Note the difference in the scale of the $y$-axis between the two subfigures.
    }
    \label{fig35}
\end{figure}
\begin{figure*}
    \centering
    \captionsetup[subfigure]{justification=centering}
    \subfloat[$n=1$, $k_z = 0.05 \micrometro$]
    {\includegraphics[width=0.22\textwidth,trim= 0.5in 0.2in 0.1in 0.0in, clip=true]{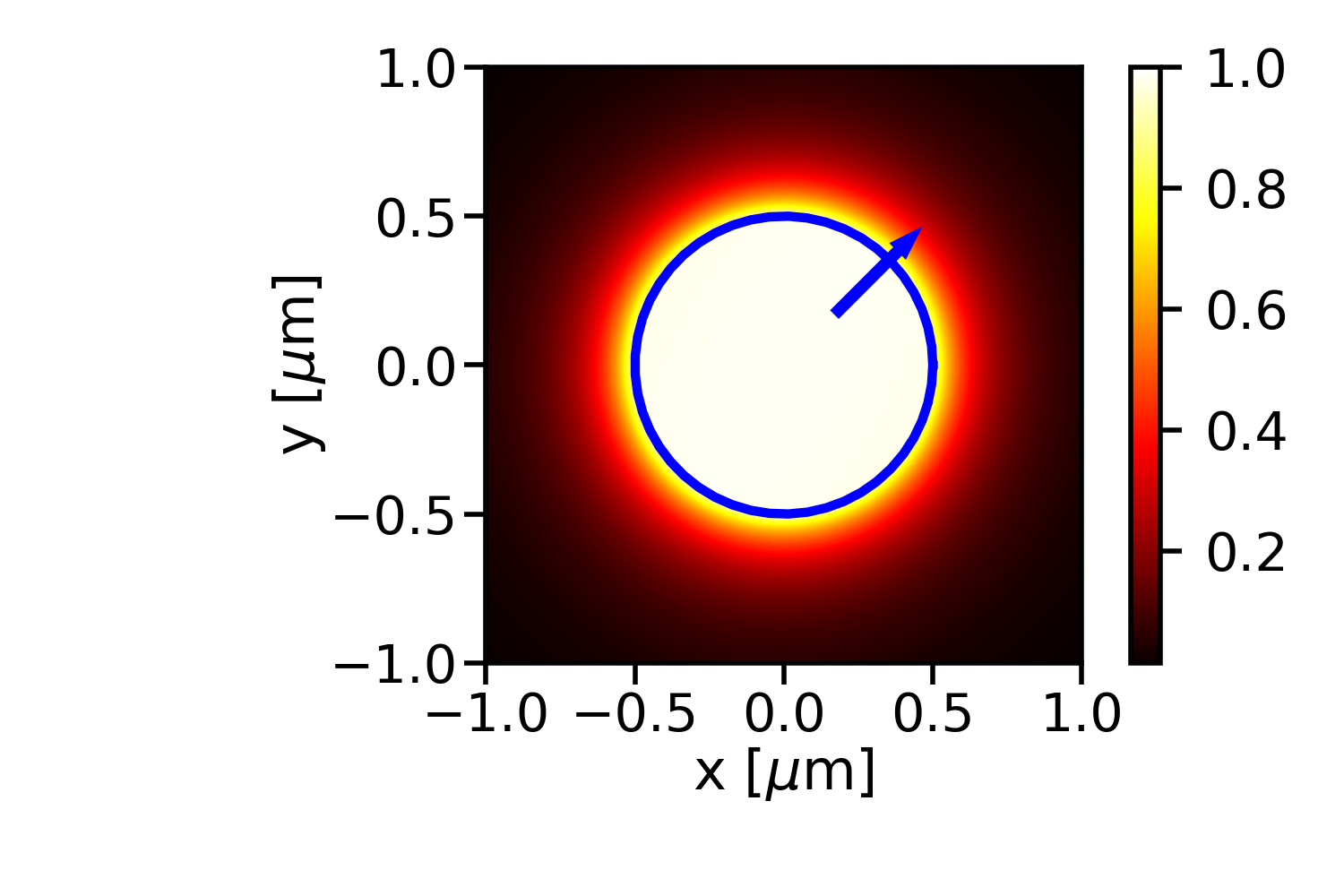}}
    \captionsetup[subfigure]{justification=centering}
    \subfloat[$n=1$, $k_z = 0.13 \micrometro$]
    {\includegraphics[width=0.22\textwidth,trim= 0.5in 0.2in 0.1in 0.0in, clip=true]{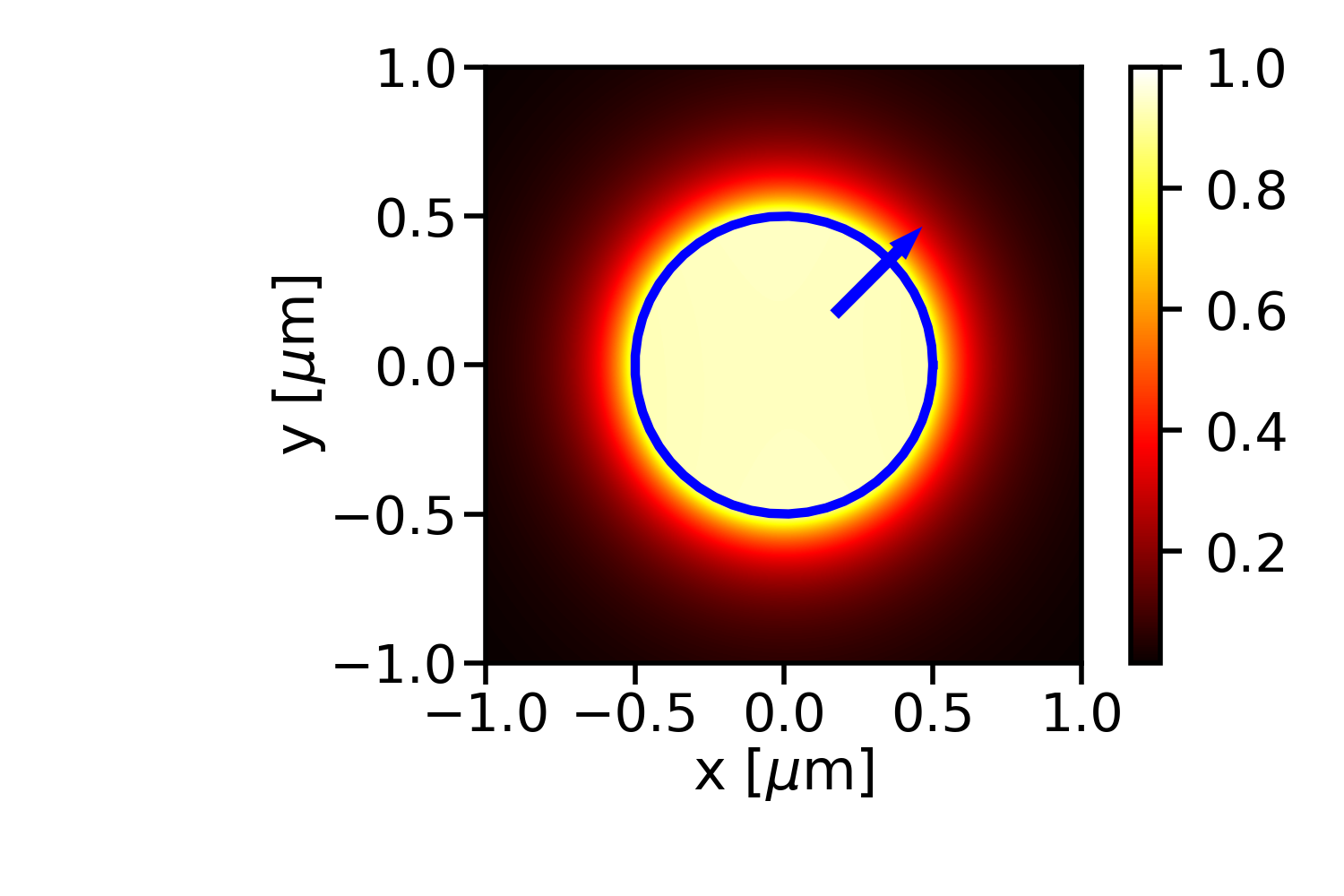}}  
    \captionsetup[subfigure]{justification=centering}
    \subfloat[$n=1$, $k_z = 0.14 \micrometro$]
    {\includegraphics[width=0.22\textwidth,trim= 0.5in 0.2in 0.1in 0.0in, clip=true]{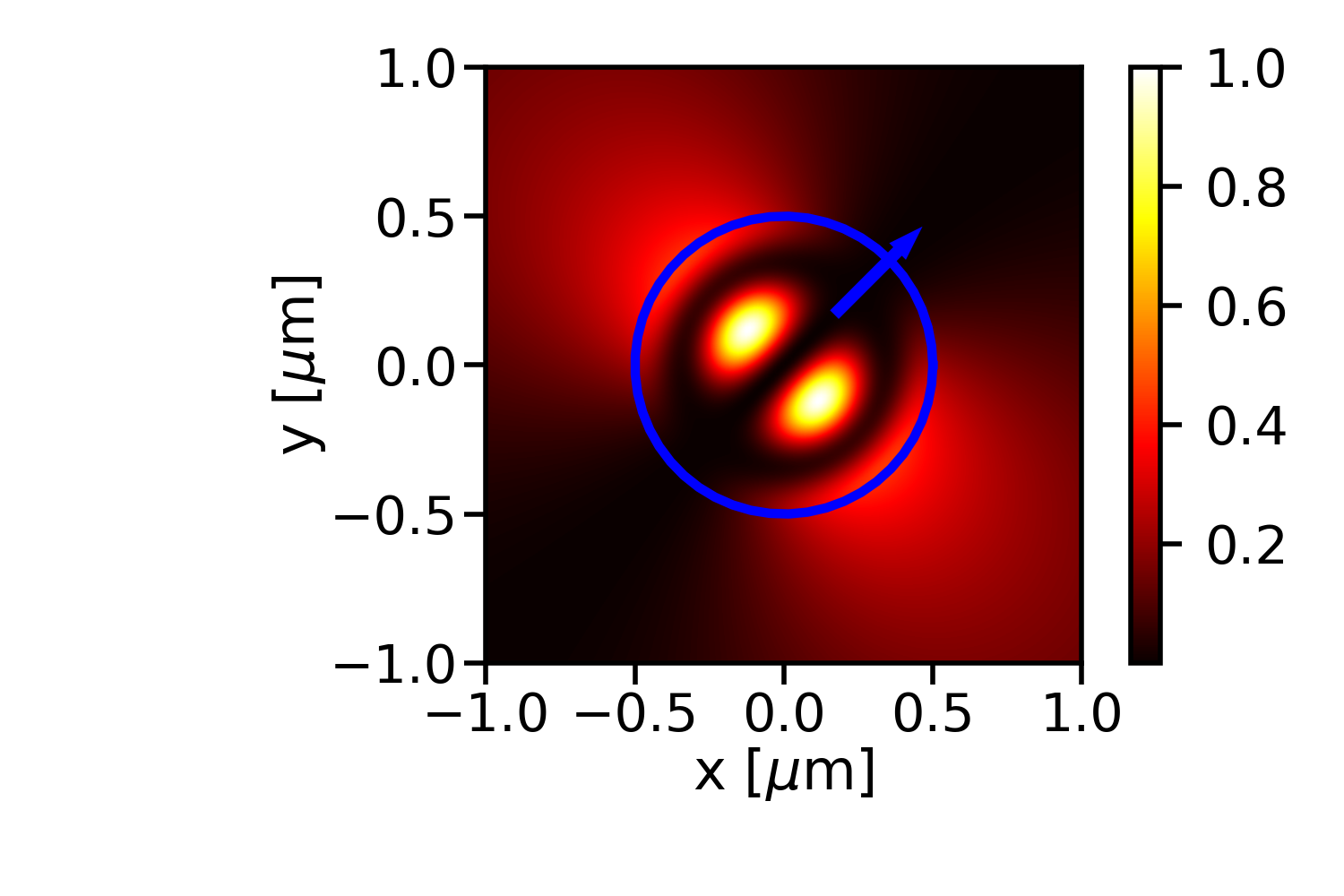}}
    \captionsetup[subfigure]{justification=centering}
    \subfloat[$n=1$, $k_z = 0.5 \micrometro$]
    {\includegraphics[width=0.21\textwidth,trim= 0.0in 0.0in 0.0in 0.0in, clip=true]{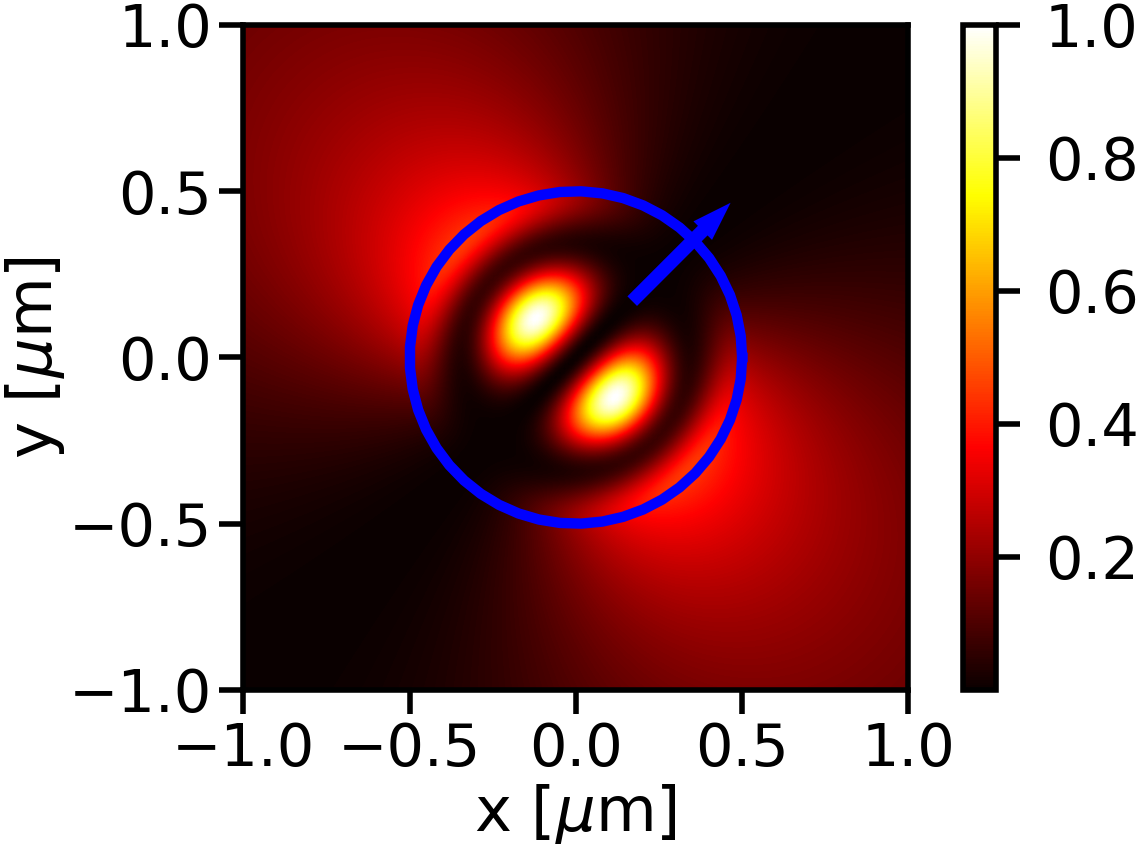}}
    \\
    \subfloat[$n=2$, $k_z = 0.05 \micrometro$]
    {\includegraphics[width=0.22\textwidth,trim= 0.5in 0.2in 0.1in 0.0in, clip=true]{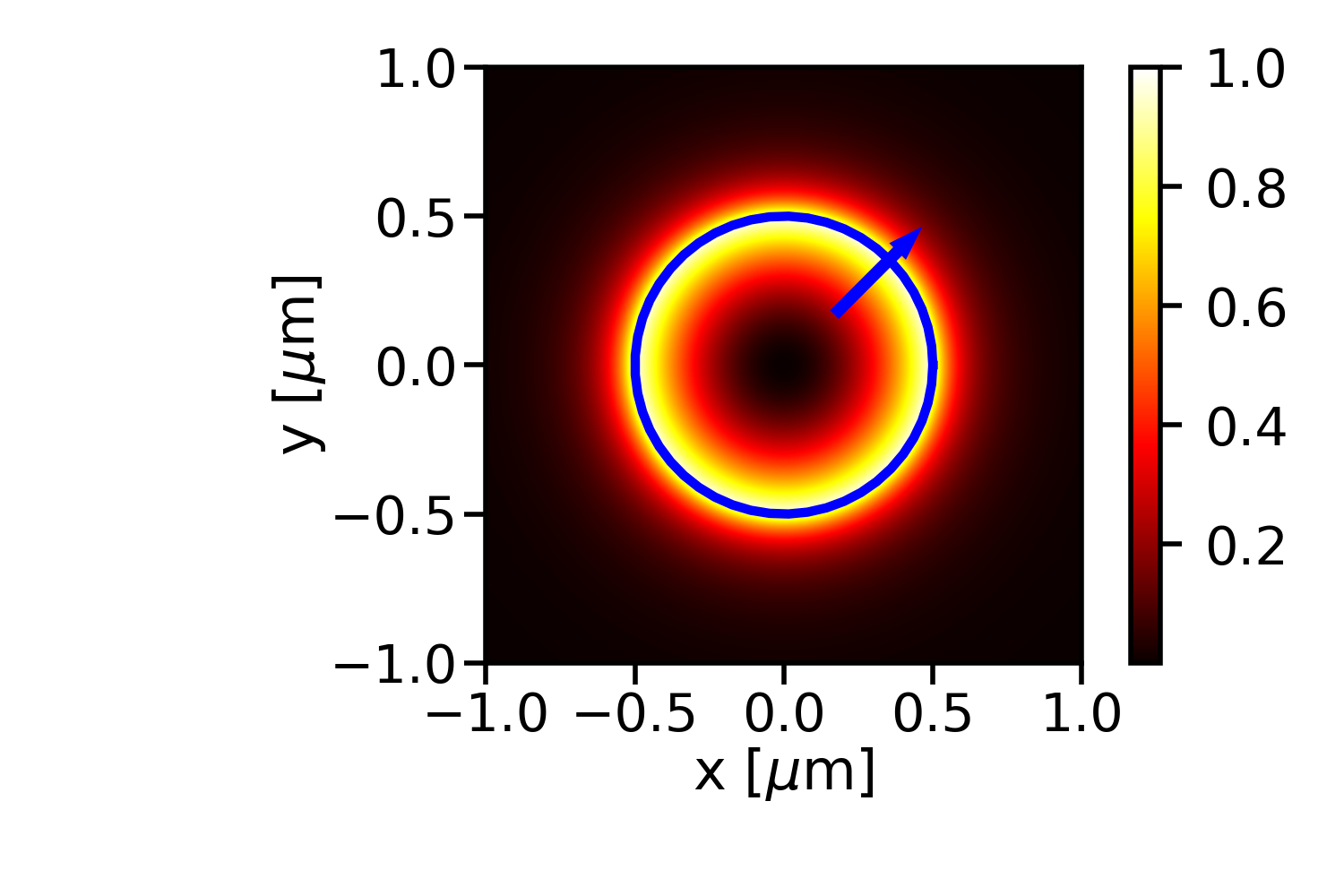}}
    \captionsetup[subfigure]{justification=centering}
    \subfloat[$n=2$, $k_z = 0.13 \micrometro$]
    {\includegraphics[width=0.22\textwidth,trim= 0.5in 0.2in 0.1in 0.0in, clip=true]{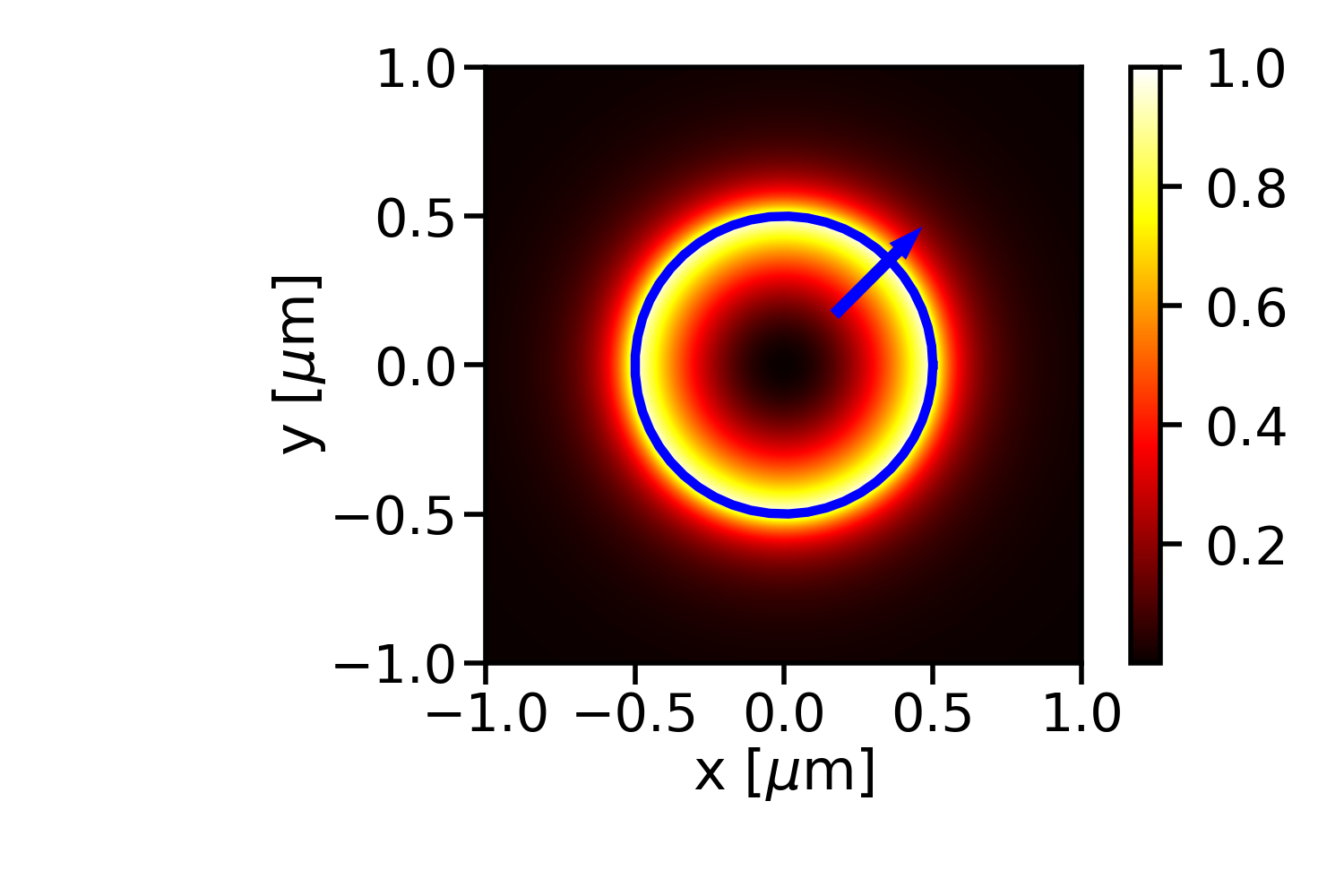}}  
    \captionsetup[subfigure]{justification=centering}
    \subfloat[$n=2$, $k_z = 0.14 \micrometro$]
    {\includegraphics[width=0.22\textwidth,trim= 0.5in 0.2in 0.1in 0.0in, clip=true]{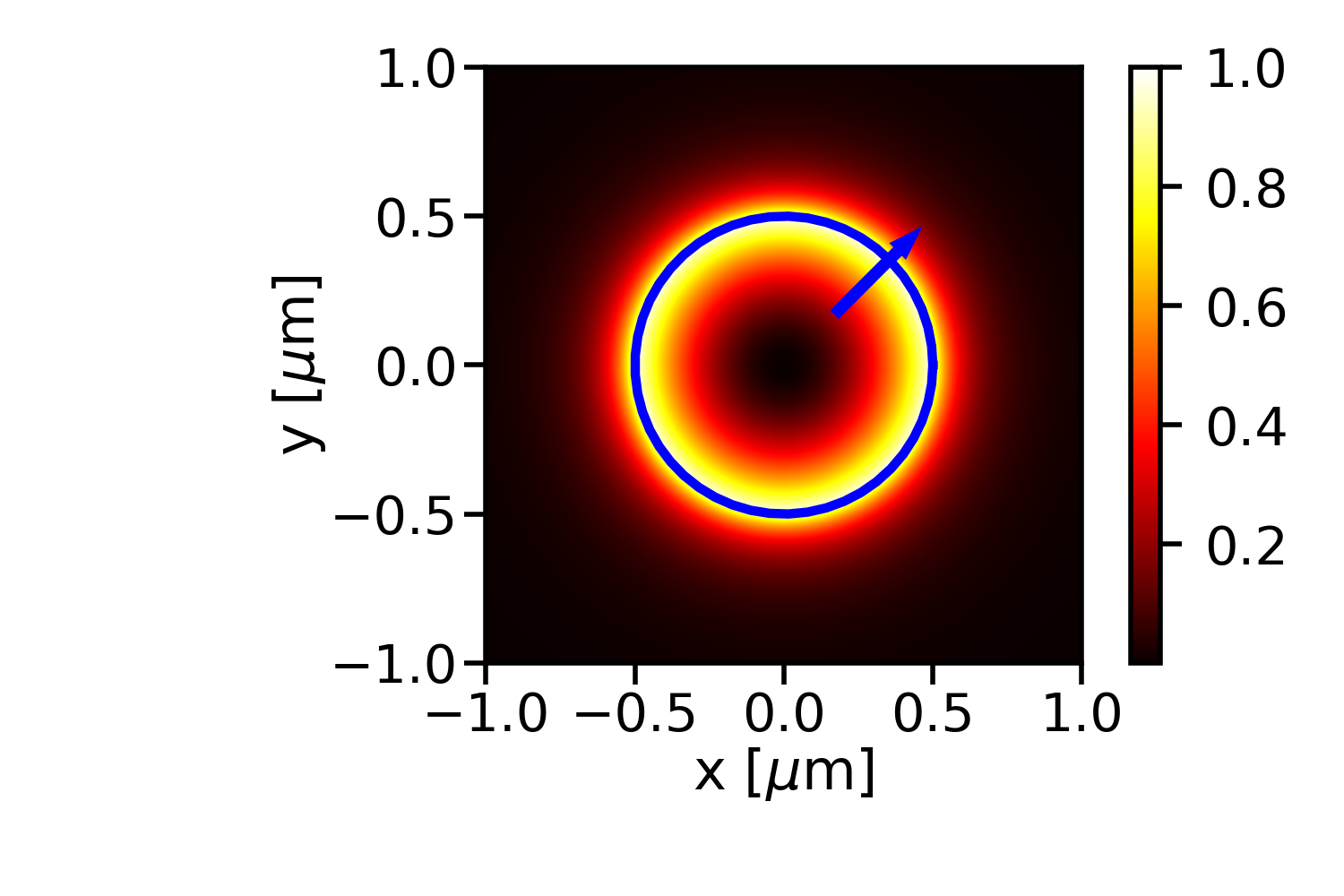}}  
    \captionsetup[subfigure]{justification=centering}
    \subfloat[$n=2$, $k_z = 0.5 \micrometro$]
    {\includegraphics[width=0.21\textwidth,trim= 0.0in 0.0in 0.0in 0.0in, clip=true]{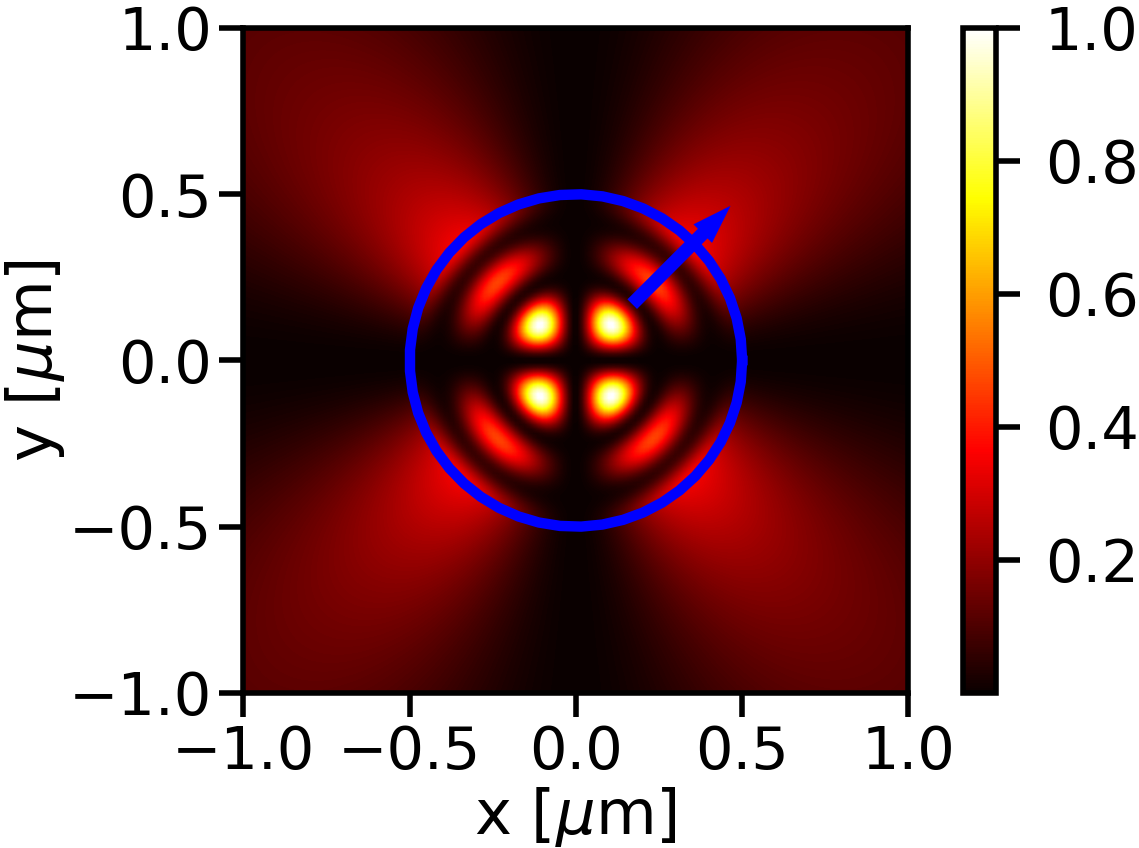}}
    \caption
    {$|\mathbf{E}|^2$ field for the excitation with a point-like dipole at $\rho_D = 0.25$, $\mu m$, $\varphi_D = \pi/4$, $z_D = 0 \mu m$, $\mu_c = 0.3 eV$, $p_x = 1$, $p_y = 1$, $p_z = 1$, $A_o = B_o = 1$. The rest of parameters are the same as in Fig. \ref{fig3}. The values of $\omega$ and $\Im (\varepsilon_1)$ are very close to the lasing condition of each case. Plots are normalized to the maximum value of $|\mathbf{E}|^2$.}
    \label{fig4}
\end{figure*}

The permittivity of the active medium $\varepsilon_1$ is a frequency-dependent property in general. However, when the response of the system is approximately the same for the frequency interval of interest, the wideband approximation can safely be used. This approximation, which implies taking $\varepsilon_1$ simply as a complex number, is useful for a first exploration of the system as it allows us  to calculate the optical response of a system independently of the characteristics of the active medium. As shown for example in Refs. \cite{arnold2015,passarelli2019}, under the appropriate conditions, this approximation provides a reliable first approach to the \ra{gain-loss compensation condition}.
\ra{Accordingly, we use this approximation in Section \ref{sec:results}\ref{sec:general} for a general analysis of the lasing conditions, while in Section \ref{sec:results}\ref{sec:realistic} we discuss a more realistic approach to the gain medium.}

\section{Results} \label{sec:results}
\subsection{\ra{Lasing conditions: General analysis using the wideband approximation.}} \label{sec:general}
\ra{In this section, we provide an analysis of the general behavior of the lasing conditions for the first multipolar modes. For this purpose, a wideband approach for the active medium is ideal since it avoids the complexity of the frequency dependence of the media.
The main objective of this section is to numerically solve the fully retarded dispersion relation Eq. (\ref{eq:disp01}).} 
The values of $\omega$ and $\Im (\varepsilon_1)$ which exactly compensate plasmon losses are the solutions of the dispersion relation for real eigenfrequencies and are called critical values. We will denote them as $[\omega]_c$ and $[\Im (\varepsilon_1)]_c$ respectively.
They provide a criterion to assess the lasing conditions, i.e., the resonant frequency and minimum value of gain that is required to significantly enhance electromagnetic fields.

Figs. \ref{fig3} shows the critical values of gain, $[\Im (\varepsilon_1)]_c$, as a function of the chemical potential of graphene, $\mu_c$, for different values of $k_z$. Let us take first the case of the dipolar mode, $n=1$, at low $k_z$ values (Figs. \ref{fig3a} and \ref{fig3b}).
We can see that although there is a small increase in the values of $[\Im (\varepsilon_1)]_c$ when $k_z$ increases, the general shape of the curves is the same.
This behavior is true but only up to a critical value of $k_z$.
Note that in Figs. \ref{fig3b} and \ref{fig3c} as $k_z$ increases from $0.13\micrometro$ to $0.14\micrometro$, the critical gain profile radically changes, becoming almost constant for $0.14\micrometro$ (note the scale of the figure). Importantly, there is an increase of more than one order of magnitude in the values of $[\Im (\varepsilon_1)]_c$. A similar behavior is also observed for the other modes but at much higher values of $k_z$, compare Fig. \ref{fig3c} with \ref{fig3d}.
\rad{
We have checked that in these examples, the behavior with $\mu_c$ of the ratio $R/\lambda$ between cylinder radius and wavelength is similar to that of the real part of $k_t$  shown in Fig. 4. For example, for the first mode and $k_z = 0.13 \micrometro$, $R/\lambda$ vs $\mu_c$ behaves as $k_t$  in Fig. 4a, reaching values between 0.012 and 0.018, whereas for $k_z = 0.13 \micrometro$, those values are between 0.441 and 0.442. }

As discussed in several works, see for example Ref. \cite{passarelli2019}, the gain-loss compensation condition depends on several factors. Basically, the total gain of a given mode can be evaluated by the integral of $|E|^2$ over the volume of the active medium times the imaginary part of the dielectric constant of that region. Optical losses have two contributions: Ohmic and radiation losses.
Radiation losses are due to the radiation being emitted by the system and is expected to have a strong dependence with $k_z$, where a given mode should shift from being radiative (large losses at low values of $k_z$) to nonradiative (small losses at high values of $k_z$).

The scattering coefficient seems to be a good indicator of the radiative/nonradiative character of a mode since it is proportional to the scattered power. 
However, at precisely the gain-loss compensation condition, the coefficients $B_n$ and $D_n$ in Eq. \ref{eq:Q_scat} diverge for a given mode and $k_z$ value. This causes the scattering coefficient to also diverge, which makes its use cumbersome.
Despite this, for the same values of the coefficients $B_n$ and $D_n$, $k_{t,2}$ (and to a lesser extent $\omega$ since it does not change significantly) is what determines the radiative character of a mode. Note also in Eq. \ref{eq:k_t} how $k_{t,2}$ and $k_z$ are related. Following this line of thought, we will associate larger values of $k_{t,2}$ with less radiative modes.

Figure \ref{fig35} shows the value of $k_{t,2}$ for the dipolar mode as a function of $\mu$ for $k_z=0.13\micrometro$ and $k_z=0.14\micrometro$.
We can observe that there is a jump of about two orders of magnitude in the values of $k_{t,2}$, plus an abrupt change in the shape of the function.
This indicates that the mode has shifted from being radiative (for $k_z=0.13\micrometro$) to an essentially nonradiative mode (for $k_z=0.14\micrometro$).
A similar behavior is observed for the other modes but at higher values of $k_z$ (not shown in the figure).
The shift of the character of the mode seems to contradict the results of Fig. \ref{fig3} which clearly shows that, for the dipolar mode at $k_z=0.14\micrometro$, more gain is needed to compensate for losses.
The solution to this apparent contradiction can be found in Fig. \ref{fig4} where we show the value of $|E|^2$ for the lasing condition of the different modes and for different values of $k_z$.
We can observe that when the dipolar mode becomes nonradiative ($k_z=0.14\micrometro$), the shape of  the mode radically changes and there is a considerable reduction of the electric field inside the \ra{active region}. 
\footnote{\ra{Gain-loss compensation condition implies that the absorbed power equals the scattered power, $W_\mathrm{abs}$ and $W_\mathrm{sca}$ respectively, where
$W_\mathrm{abs} \propto \sum_\alpha \mathrm{Im} (\varepsilon_\alpha) \int_{V_\alpha} k |E|^2 dV$, being $V_\alpha$ (with $\alpha=\mathrm{act},\mathrm{pas}$) the volume over the active or passive regions \cite{passarelli2019}.}}
Therefore, it is reasonable that more gain is now necessary to compensate for losses. Clearly, this effect is dominating over the reduction of radiative losses. 
Fig. \ref{fig4} also shows that 
a similar transition also occurs for the quadrupolar mode, but for a much higher value of $k_z$. The rest of the studied modes follow a similar change in the pattern of $|E|^2$.

\rad{We have analyzed the origin of the change in the field distribution when the value of $k_z$ increases. We found that there is a switch in the character of the modes which pass from being plasmonic (at low $k_z$ values) to being photonic (at high $k_z$ values). This was confirmed after comparing the lasing frequency obtained by our numerical procedure with the laser frequency predicted by a quasi-static approximation of the dispersion relation of plasmonic modes in extended graphene, see Appendix \ref{app:aprox}. We observed that the frequency obtained numerically coincides reasonably well with that predicted by \eqref{b5} but only up to the critical value of $k_z$, where there is a shift in the field distribution. From that point, there is a strong disagreement between both quantities. Furthermore, for plasmonic modes (low values of $k_z$) we always observed a strong dependence with $\mu_c$ in  both the frequency and the gain required for lasing, whereas photonic modes are insensitive to $\mu_c$, see Figs. 3 and 6.}
\begin{figure}[t]
    \centering
    \subfloat[$n=2$ (Fig. 5g)]{\includegraphics[width=0.24\textwidth
    ]{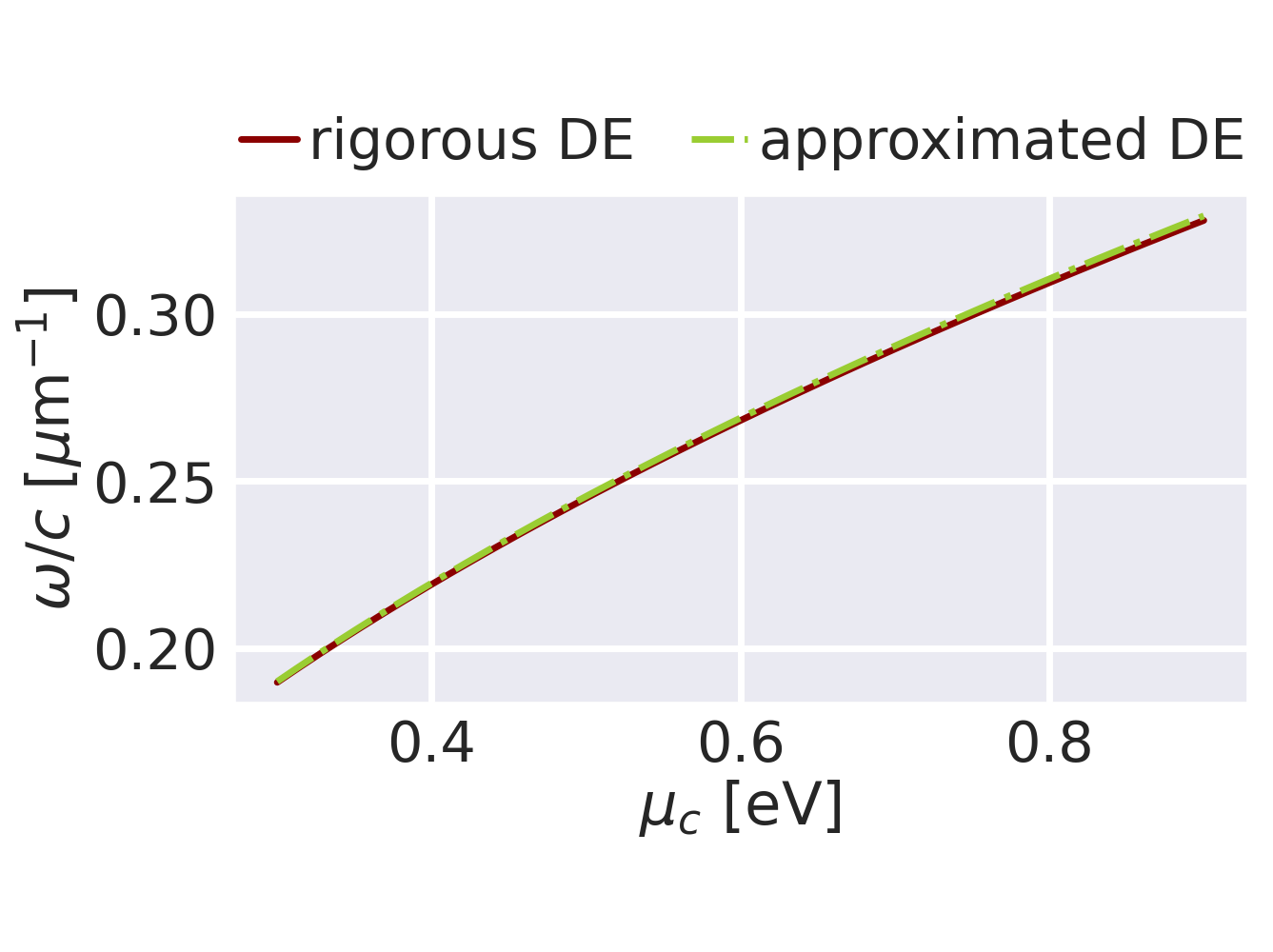}\label{nuevaa}}
    \subfloat[$n=1$ (Fig. 5c)]{\includegraphics[width=0.24\textwidth
    ]{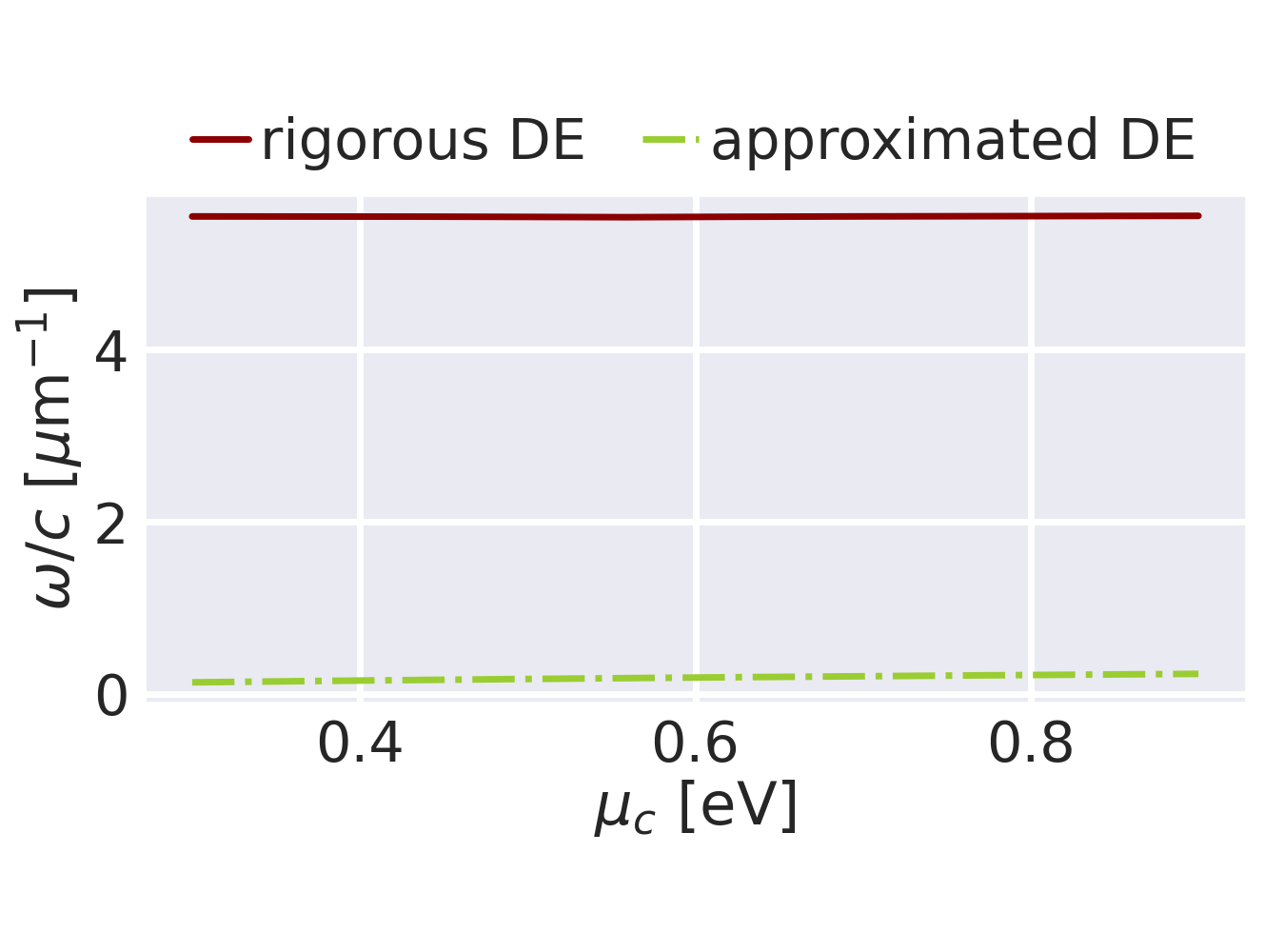}\label{nuevab}}
    \caption{Lasing frequency vs $\mu_c$ for $k_z = 0.14 \micrometro$ and for the spatial field distributions shown in Fig 5g and 5c. The red curve corresponds to values obtained finding the the roots of the determinant of $\mathbb{M}_n$ in \eqref{eq:disp01} and the green curve to values obtained the approximated dispersion equation \eqref{b5}. }    \label{fig:nueva}
\end{figure}

\subsection{A realistic example for a lasing device}\label{sec:realistic}

Let us consider doped silicon as the core of our device. When silicon is doped with group-V donors, it can provide intra-center transitions that are Raman active from $ 1-7$THz \cite{pavlov2013physical}. The lattice absorption losses are low in this range and the thermal conductivity of the material is large, which should help to prevent the overheating of the device.The cross-sections of the transitions are also extensive.

\ra{Recent experimental advances allow for the covering of silicon nanowires with graphene, which encouraged us to study the present geometry. The fabrication of this type of architecture  was boosted by its suitability for lithium batteries and hydrogen production \cite{wang2013adaptable,cho2011nitrogen,sim2015n}. In the cited references the Si nanowires were produced by metal etching, and the carbon coating due to depositing graphene sheets \cite{wang2013adaptable,cho2011nitrogen} or chemical vapor \cite{sim2015n}. 
}

\begin{figure}[h!]
\centering
\subfloat[$\mu_c$ = 0.24 eV, $k_z = 0.32\micrometro$]{\includegraphics[width=0.24\textwidth]{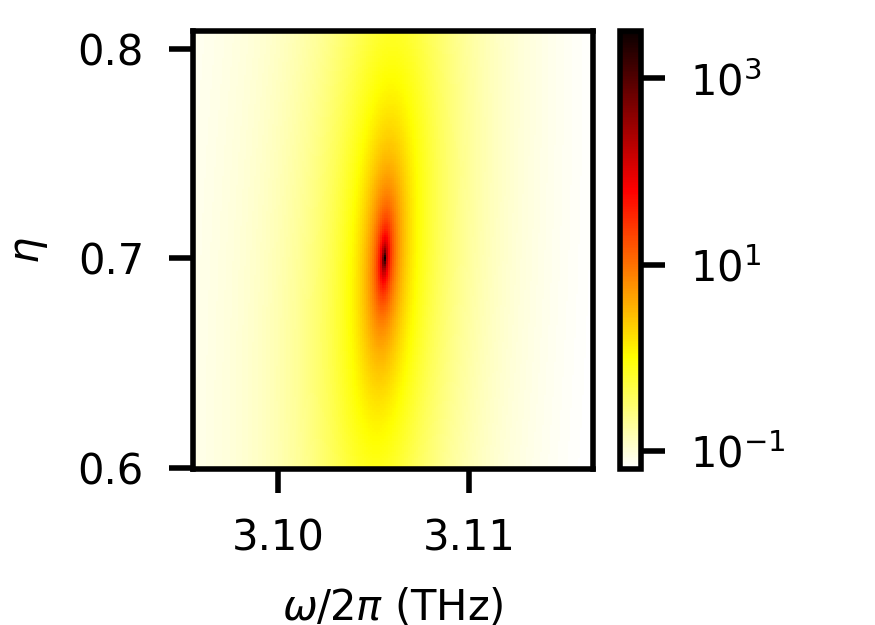}\label{fig6 para freq 3.15}}
\subfloat[$\mu_c$ = 0.4 eV, $k_z = 2.16\micrometro$]{\includegraphics[width=0.24\textwidth]{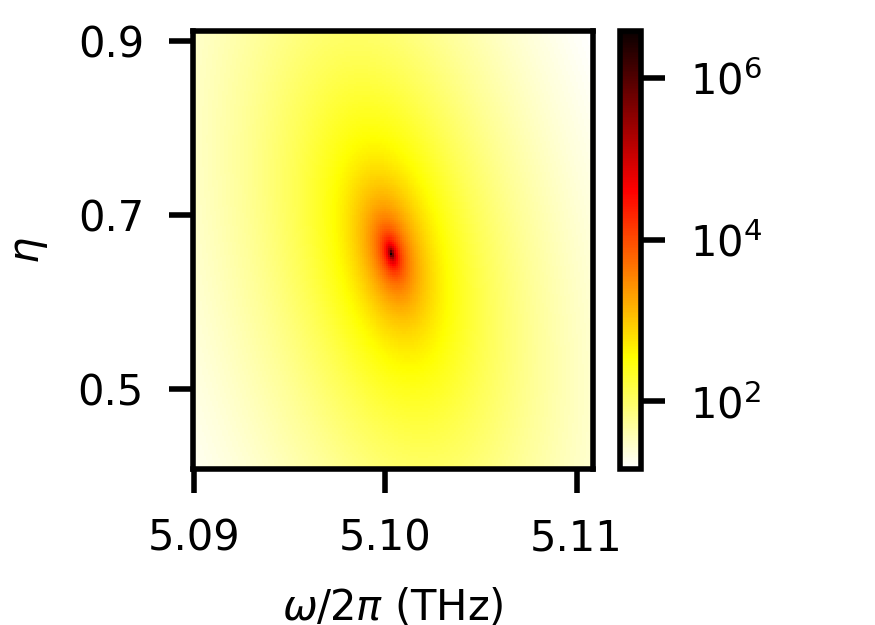}\label{fig6 para freq 5.09}}
\caption{Scattering coefficient as a function of $\eta$ and $\omega$. Parameters not indicated in the figures are the same as Fig. \ref{fig3} except for $\varepsilon_1$ which is taken from Fig. \ref{fig5}.
}
\label{fig6}
\end{figure}

Using the obtained values of $\varepsilon_1$\nico{, shown in Fig. \ref{fig5}}, we calculated the scattering coefficient (Eq. \ref{eq:Q_scat}) as function of $\eta$ and $\omega$ for a fixed value of $k_z$, see Fig. \ref{fig6}. In this case, singularities of the scattering coefficient mark the lasing conditions, see Ref. \cite{prelat2021,passarelli2019,passarelli2016}. As the figure shows, the lasing of the system, even at higher values of $k_z$, is achieved at relatively low values of population inversion, $\eta$.
\begin{figure}[h!]
\centering
\subfloat[$\mu_c$ = 0.24 eV]{\includegraphics[width=0.24\textwidth]{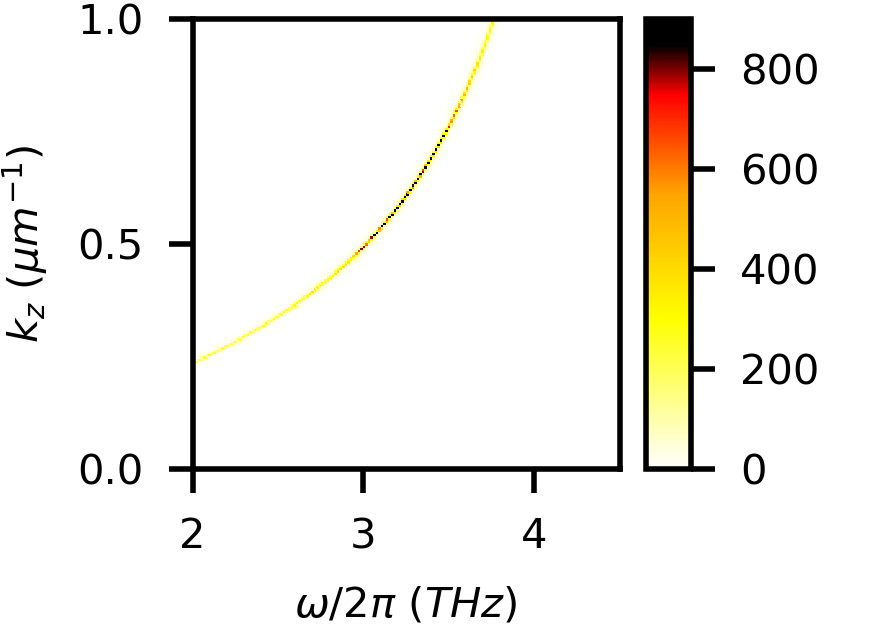}\label{fig7a}}
\subfloat[$\mu_c$ = 0.63 eV]{\includegraphics[width=0.24\textwidth]{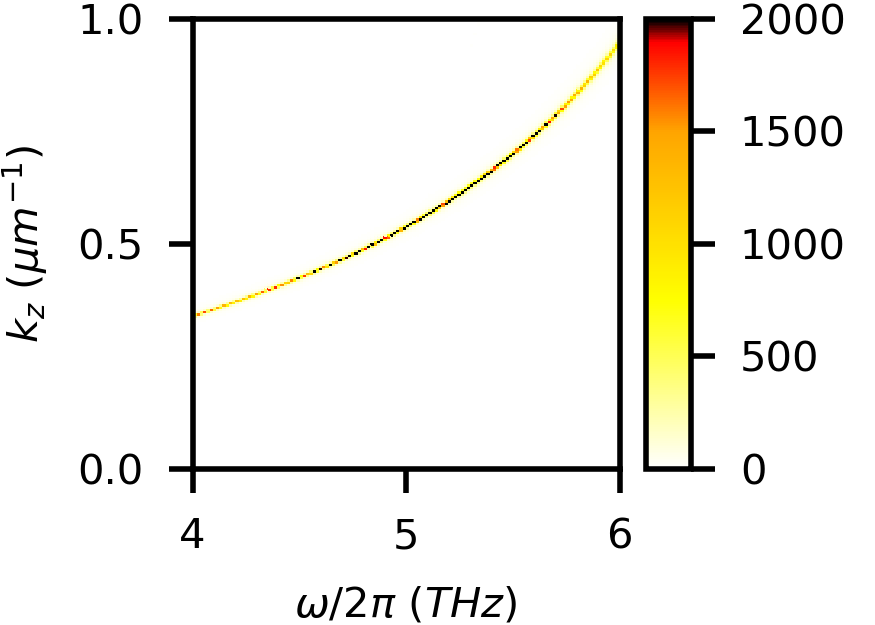}\label{fig7b}}
\caption{Scattering coefficient as a function of $k_z$ and $\mu_c$.  In both plots we used $\eta = 0.92$, which maximized the scattering for $k_z=0.57\micrometro$ and $\omega/2\pi=3.17$THz (for $a$) or $5.11$THz (for $b$). Parameters not indicated in the figures are the same as Fig. \ref{fig6}.
}
\label{fig7}
\end{figure}

In practice, it might be difficult to experimentally select a particular value of $k_z$. 
\footnote{In principle, it is possible to externally control $\mu_c$ (electrically) and $\eta$ (with the power of the pumping laser), which would allow to indirectly select at least some modes within a specific range of $k_z$ values, while others are kept below the lasing condition.}
For that reason, we show in Figures \ref{fig7a} and \ref{fig7b} a more realistic situation than Fig. \ref{fig6}, where the value of $\eta$ is fixed (instead of $k_z$). 
The figures show that by simply changing the chemical potential of graphene it is possible to select which of the two types of transitions of the active medium is being used for lasing (population inversion at $5.09$THz or stimulated Raman scattering at $3.15$TH).

\section{Conclusions}\label{sec:conclusions}

We have presented an analytical method for finding the lasing conditions and the electromagnetic characteristics of a wire coated with graphene.
For the material inside the cylindrical wire, we used two different models: i) an active medium with a frequency-independent imaginary part and ii) a realistic model of active medium which includes a Lorentzian dependence with the frequency, the possibility of absortion of radiation, stimulated emission of radiation, and stimulated raman scattering. We have used the first model to characterize in general terms the behavior of the system while the second one has been used to assess the feasibility of our proposal.

Our results have shown that graphene-coated nanocylinders are able to sustain the stimulating emission of propagating surface plasmon polaritons in the terahertz range of frequency.
The values required for population inversion seem reasonably low and should be accessible experimentally.
Moreover, the system has proved to be highly tunable through the electric control of the graphene chemical potential.
\ra{This may allow the use of current semiconductor technology to quickly and easily control the frequency of micro/nanometer-sized terahertz radiation sources.}

A detailed analysis of the stimulated emission of different modes has shown that there is an abrupt change in the gain required for lasing when the modes transition from being radiative to nonradiative. Counterintuitively, nonradiative modes require more gain compensation. This odd behavior is explained in terms of a change in the field distribution inside the gain medium, a change that occurs when the character of the modes shifts from plasmonic to photonic.

\section*{Acknowledgments} 
We acknowledge financial support by Consejo Nacional de Investigaciones Cient\'ificas y T\'ecnicas (CONICET); Secretar\'ia de Ciencia y Tecnolog\'ia de la Universidad Nacional de C\'ordoba (SECYT-UNC); and Agencia Nacional de Promoci\'on Cient\'ifica y Tecnol\'ogica (ANPCyT, PICT-2018-03587). LP and RD are grateful to Dr. Mauro Cuevas for useful discussions at an early stage of this research work.

\appendix

\section{Boundary conditions}          \label{app:boundary}
\setcounter{equation}{0}
\renewcommand{\theequation}{A{\arabic{equation}}}

At the boundary $\rho = R$ (graphene layer) 
the tangential components of the electric field are continuous and the tangential components of the magnetic field have a discontinuity proportional to the graphene surface current \cite{kubo2} 
\begin{align}
& \hat{\rho} \times (\vec{E}^{(1)} - \vec{E}^{(2)}) \Big|_{\rho=R}  = \vec{0},  \\
& \hat{\rho} \times (\vec{H}^{(1)} - \vec{H}^{(2)}) \Big|_{\rho=R}  = \dfrac{4\pi}{c} \sigma(\omega) (E_z \hat{z} + E_\varphi\hat{\varphi} )  \Big|_{\rho=R}.
\end{align}
\rad{
Adding the contribution of the sources (plane wave or dipole) to the electromagnetic fields given by Eqs. (\ref{Ez12p}-\ref{Ht12p}) and using the boundary conditions, the following system of linear equations 
\begin{equation*}
\mathbb{M}_n\vec{X}_n = \vec{V}_n,
\end{equation*}
is obtained for the unknown amplitudes 
\begin{eqnarray*}
\vec{X}_n=\left[A_n, \, B_n, \, C_n, \, D_n \right]^T .
\end{eqnarray*}
}
The matrix of this system is given by 
\begin{align*}
 \mathbb{M}_n = 
\begin{psmallmatrix}
J_1 & - H_2  & 0 & 0\\
\\
\dfrac{4\pi \sigma(\omega) }{c}  J_1 + \dfrac{i\omega\varepsilon_1}{c {k}_{t,1}} J'_1 & -\dfrac{i\omega\varepsilon_2}{{k}_{t,2}c} H'_2 & -\dfrac{k_z n}{{k}^2_{t,1}R}   J_1  & \dfrac{k_{z} n}{{k}^2_{t,2}R} H_2 \\
\dfrac{4\pi \sigma(\omega) k_z n}{cR{k}^2_{t,1}}  J_1 & 0  & J_1 +\dfrac{4\pi i\sigma(\omega) \omega \mu_1}{c^2 {k}_{t,1}} J'_1 & -  H_2 \\
-\dfrac{k_z n}{R {k}^2_{t,1}}  J_1  & \dfrac{k_z n}{R {k}^2_{t,2}}  H_2  &  - \dfrac{i\omega \mu_1}{c{k}_{t,1}} J'_1 & \dfrac{i\omega \mu_2}{c{k}_{t,2}} H'_2
\end{psmallmatrix} 
\end{align*}
\noindent \textnormal{with } 
$J_m = J_n(k_{t,m}R)$, $H_m = H^{(1)}_n(k_{t,m}R)$, $J'_m = \dfrac{\partial J_n(k_{t,m}\rho)}{\partial (k_{t,m}\rho)}\Big|_{\rho=R}$ and $H'_m = \dfrac{\partial H^{(1)}_n(k_{t,m}\rho)}{\partial (k_{t,m}\rho)}\Big|_{\rho=R}\,$. The prime denotes the first derivative with respect to the argument of the function. 

\rad{
For the eigenmode problem, $\vec{V}_n=0$. For the plane wave excitation problem, $\vec{V}_n$ takes the form
\begin{align}
&\vec{V}_n=i^n\Big[B_o  J_2, \, -\dfrac{ k_z n }{R {k}^2_{t,2}}A_o  J_{2} + \dfrac{i\omega \varepsilon_2 }{c {k}_{t,2}} B_o  J'_{2}, \nonumber\\
& A_o J_2, \, - \dfrac{i\omega \mu_2 }{c {k}_{t,2}} A_o  J'_2 -\dfrac{ k_z n }{R {k}^2_{t,2}} B_o  J_2\Big]^T,
\end{align}
and when the system is excited by an electric dipole $\vec{p}$ located inside the cylinder at the position $\vec{r}_D = \rho_D\hat{\rho} + \varphi_D \hat{\varphi} $  ($\rho_D<R$) from the center of the cylindrical wire, $\vec{V}_n$ takes the form \cite{novotny2006, Cuevas2017_dipolo_teoria}
%
\begin{align}
\vec V_n= 
 \begin{bmatrix}
- {e_{z n}(R)}\\
\\
 \dfrac{k_z  n}{R} \dfrac{ {h_{z n}(R)}}{k^2_{t,1}}  - i  \dfrac{\varepsilon_1  {e'_{z n}(R)} \omega }{k^2_{t,1}c} - \dfrac{4\pi \sigma}{c} {e_{z n}(R)} \\
\\
- {h_{z n}(R)} - \dfrac{4\pi \sigma}{c} \dfrac{1}{k^2_{t,1}}\left[ \dfrac{k_z n }{R} {e_{z n}(R)} + \dfrac{i\mu_1 \omega}{c}  {h'_{z n}(R)} \right] \\
\\
\dfrac{i\mu_1 c}{\omega}  {h'_{z n}(R)} + \dfrac{k_z n }{R } \dfrac{ {e_{z n}(R)}}{k^2_{t,1}} 
 \end{bmatrix}
\end{align}
where 
\begin{align}
e_{z  n}(R) &= \dfrac{k_{t,1}}{4\varepsilon_1} e^{-in\theta_D}  H^{(1)}_n(k_{t,1}\rho) \Big[2ip_z J_n \left\{ \dfrac{n(n+1)}{\rho^2  k_{t,1}} - k_{t,1} \right\}   \nonumber\\[3mm]
& - k_z p_+ e^{-i\theta_D}J_{n+1} +  k_z p_- e^{i\theta_D}J_{n-1}  \Big]    \,, 
\end{align}
\begin{align}
h_{z  n}(R) &=- \dfrac{i\omega}{4c} k_{t,1} e^{-in\theta_D} H^{(1)}_n \Big\{ p_+ \,e^{-i\theta_D} J_{n+1}   \nonumber\\[3mm]
& + p_- \,e^{i\theta_D} J_{n-1} \Big\}   \,,  
\end{align}
the argument of the special functions is $k_{t,1}\rho_D$ and $p_+ = p_x + i p_y , \,    p_- = p_x - i p_y,$. 
}

\section{Approximated dispersion equation for plasmonic modes}          \label{app:aprox}
\setcounter{equation}{0}
\renewcommand{\theequation}{B{\arabic{equation}}}

\rad{
We assume the existence of a surface plasmon propagating along the cylindrical graphene surface (Fig. \ref{dibujo}) with wavevector 
\begin{equation} 
\vec{k}_{S P}=k_{\varphi} \hat{\varphi}+k_{z} \hat{z}. \label{b1}
\end{equation}
Due to the rotational symmetry of the system, a mode will form when an integral number ($n$) of azimuthal wavelengths fits around the graphene cylinder
\begin{equation} 
k_{\varphi} 2 \pi R=2 \pi n. \label{b2}
\end{equation}
Assuming the validity of the quasistatic approximation (wire radius small compared to the modal wavelength, $R \ll \lambda_n = \Re{2\pi c / \omega_n}$) and that the magnitude of the plasmon wavevector can be approximated by the value corresponding to a planar graphene interface \cite{maximo1,jablan}, we have  
\begin{equation} 
k_{S P}=i \frac{\omega}{c} \frac{\varepsilon_{1}+\varepsilon_{2}}{4 \pi \sigma / c}. \label{b3}
\end{equation}
Taking into account that 
\begin{equation} 
k_{S P}^{2} =k_{\varphi}^{2} + k_{z}^{2} \,, \label{b4}
\end{equation}
we get the following implicit dispersion equation for $\omega_n(k_z)$
\begin{equation} 
\frac{\omega_n}{c}=\sqrt{\frac{-\left[k_{z}^{2}+\left(\frac{n}{R}\right)^{2}\right]}{\left[\frac{\varepsilon_{1}+\varepsilon_{2}}{4 \pi \sigma\left(\omega_n\right) / c}\right]^{2}}}.  \label{b5}
\end{equation}
Please note that, in the spectral region of interest $\sigma$, given by \eqref{intra} and \eqref{inter} is almost a purely imaginary number, with $\Im{\sigma}>0$ and a very small real part \cite{kubo1,kubo2}. }

\section*{Disclosures}
The authors declare no conflicts of interest.

\section*{Data availability}
\rad{Data underlying the results presented in this paper are not publicly available at this time but may be obtained from the authors upon reasonable request.}

\medskip 
\bibliography{cites-08}

\end{document}